\documentclass[review]{elsarticle}
\usepackage[top=0.5in, bottom=1in, left=1in, right=1in]{geometry}

\usepackage{amsmath} 
\usepackage[colorlinks,linkcolor=blue,anchorcolor=red,citecolor=green]{hyperref}

\journal{Annals of Physics}

\begin{document}

\begin{frontmatter}

\title{Progress in vacuum susceptibilities and their applications to the chiral phase transition of QCD}

\author[add1,add6]{Zhu-Fang Cui\corref{corrauthor}}
\ead{phycui@nju.edu.cn}

\author[add2,add6]{Feng-Yao Hou}

\author[add1,add3,add6]{Yuan-Mei Shi}

\author[add1,add4]{Yong-Long Wang}

\author[add1,add5,add6]{Hong-Shi Zong\corref{corrauthor}}
\cortext[corrauthor]{Corresponding author}
\ead{zonghs@nju.edu.cn}

\address[add1]{Department of Physics, Nanjing University, Nanjing 210093, China}
\address[add2]{Institute of Theoretical Physics, CAS, Beijing 100190, China}
\address[add3]{Department of Physics and electronic engineering, Nanjing Xiaozhuang University, Nanjing 211171, China}
\address[add4]{Department of Physics, School of Science, Linyi University, Linyi 276005, China}
\address[add5]{Joint Center for Particle, Nuclear Physics and Cosmology, Nanjing 210093, China}
\address[add6]{State Key Laboratory of Theoretical Physics, Institute of Theoretical Physics, CAS, Beijing, 100190, China}

\begin{abstract}
The QCD vacuum condensates and various vacuum susceptibilities are all important parameters which characterize the nonperturbative properties of the QCD vacuum. In the QCD sum rules external field formula, various QCD vacuum susceptibilities play important roles in determining the properties of hadrons. In this paper, we review the recent progress in studies of vacuum susceptibilities together with their applications to the chiral phase transition of QCD. The results of the tensor, the vector, the axial-vector, the scalar, and the pseudo-scalar vacuum susceptibilities are shown in detail in the framework of Dyson-Schwinger equations.
\end{abstract}


\end{frontmatter}

\newpage
\tableofcontents
\newpage


\section{Introduction}\label{intro}
Nowadays, the commonly accepted theory that describes the strong interaction is Quantum Chromodynamics (QCD). Due to the asymptotic freedom nature of QCD, in the high energy and high momentum transfer region reliable calculations can be done using the perturbative theory. Numerous comparisons between the results from theoretical calculations and experimental measurements have shown that QCD is correct in the high energy field. However, in the low energy and low momentum transfer region the coupling constant $\alpha_s$ of QCD becomes large and running, consequently things have to be treated non-perturbatively; and thanks to the non-perturbative nature, the QCD vacuum is not trivial. People believe that the non-trivial properties of QCD vacuum are closely linked to spontaneous chiral symmetry breaking and confinement, which are the two main features of low-energy hadron physics. In order to characterize the QCD vacuum, people often introduce various vacuum condensates (such as the two-quark condensate, the gluon condensate, the mixed quark gluon condensates, and four-quark condensate, etc.) and many vacuum susceptibilities (such as the tensor vacuum susceptibility, the vector and the axial-vector vacuum susceptibilities, the scalar and the pseudo-scalar vacuum susceptibilities) phenomenologically~\cite{NPB.147.385--447,PLB.129.328--334,NPB.232.109--142,PR.127.1--97,PLB.271.223--230,PRD.45.1754}. In this paper, we review the recent progress in studies of vacuum susceptibilities together with their applications to the chiral phase transition of QCD. The results of various vacuum susceptibilities are shown in detail in the framework of Dyson-Schwinger equations (DSEs).

Metaphorically speaking, we can regard the QCD system as a black box, one way to study the QCD vacuum is to cause a perturbation by adding an external field to the system, such that we can learn something about the QCD vacuum indirectly by studying its responses to the external field.\footnote{In the experimental side, the physicists often measure the linear responses that are proportional to the perturbations, such as many susceptibilities, conductivity, etc. Therefore, in order to compare with experimental results the calculations of various linear responses are very important in the theoretical side.} In the theory of QCD, the quark propagator is the simplest Green function (two-point Green function), and its linear responses to the external fields reflect the properties of QCD vacuum directly. QCD vacuum susceptibilities are just the parameters that are associated with the linear responses of the quark propagator to the external fields, which can then characterize the non-perturbative properties of QCD vacuum. In the QCD sum rules external field formula, QCD vacuum susceptibilities play important roles in determining the hadron properties~\cite{PLB.129.328--334,NPB.232.109--142,PLB.271.223--230,PRD.45.1754}. For example, the strong and parity-violating pion-nucleon coupling depends crucially upon the vacuum susceptibilities of pion~\cite{PLB.367.21--27,PLB.576.289--296}, and the tensor vacuum susceptibility is closely related to the tensor charge of nucleon~\cite{PRL.67.552,NPB.375.527--560,PHYSREVD.52.2960,PRD.54.6897,PLB.395.307--310,PLB.438.242--247,PRC.59.3377,EPJC.17.129--135,PLB.557.33--37}. The nonlinear susceptibilities are also found to be correlated with the cumulant of baryon-number fluctuations in experiments~\cite{PhysRevD.68.034506,PhysRevD.71.114014,PhysRevD.72.054006,SCI.332.1525,PhysRevD.90.114031}. Due to the importance of QCD vacuum susceptibilities in low energy hadron physics field, many different methods and models to study various QCD vacuum susceptibilities have been utilized over the years. Take the scalar vacuum susceptibility as an example, there are Lattice QCD calculations~\cite{JHEP.09.073,PRD.84.094004,JHEP.01.138,PhysRevLett.110.082002}, calculations using the DSEs method~\cite{PRC.58.1758,PRD.77.076008,PLB.675.32--37}, the multi-flavor Schwinger model~\cite{PLB.318.531--536,PRD.54.1087}, the linear sigma model~\cite{PRC.68.035209,EPJA.16.291--297}, the (Polyakov loop extended) Nambu--Jona-Lasinio ((P)NJL) model~\cite{PRD.75.074013,EPJC.56.483--492,PhysRevD.88.114019}, the nonlocal chiral quark model~\cite{PRD.79.014008}, and the chiral perturbation theory~\cite{PPNP.67.337--342,PRD.87.016001}, etc. However, owing to the lack of a model independent expression for calculating the vacuum susceptibilities, different models or approximations often give different results, some of which may even be distinct. In this case, in order to provide a good starting point for studies of vacuum susceptibilities, the authors of Ref.~\cite{PRC.72.035202} proposed a model independent equation for calculating the vector vacuum susceptibility firstly, using the QCD sum rules external field formula. It should be pointed out that, their results are based on the linear response theory to an external field of fermions, so that it is universal, and can also be used in other quantum field theories (e.g., Quantum Electrodynamics in (2+1) dimensions, QED$_3$)~\cite{PRB.66.144501,PhysRevD.87.116008,PhysRevD.90.073013}. Based on this, people have also studied the axial-vector and the tensor vacuum susceptibilities in the framework of DSEs and under the rainbow-ladder approximation~\cite{PRC.73.035206,PLB.639.248--257}, while the authors of Ref.~\cite{PRC.79.035209} studied the scalar vacuum susceptibility beyond the rainbow-ladder approximation, as well as the pseudo-scalar vacuum susceptibility~\cite{PRC.81.032201}. Here we want to stress that the authors of Ref.~\cite{PRC.81.032201} derived a model independent result for the pseudo-scalar vacuum susceptibility using the isovector--pseudo-scalar vacuum polarization, whereas the authors of Ref.~\cite{PLB.669.327--330} obtained model independent results for the vector and axial-vector vacuum susceptibilities using the vector and axial-vector Ward-Takahashi identities (WTIs). In Ref.~\cite{FBS.48.31}, the authors calculated the tensor vacuum susceptibility employing the Ball-Chiu (BC) vertex~\cite{PRD.22.2542,PRD.22.2550}, and also studied the dressing effect of the quark-gluon vertex on the tensor vacuum susceptibility. Their results show that the tensor vacuum susceptibility obtained in the BC vertex approximation is reduced by about 10\% compared to its rainbow-ladder approximation value, which means that the dressing effect of the quark-gluon vertex is not very large in the calculation of the tensor vacuum susceptibility within the DSEs framework. These works promoted our understandings of the five kinds of vacuum susceptibilities, and up to now, only the scalar and the tenser vacuum susceptibilities are still model-dependent, to choose a more reliable model then becomes very important.

It is commonly accepted that with increasing temperature and/or quark chemical potential, strongly interacting matter will undergo a phase transition from the hadronic phase to the quark-gluon plasma (QGP) phase, which is also expected to appear in the ultra-relativistic heavy ion collisions~\cite{PoS.06.024}, and the chiral symmetry which has been broken dynamically is also supposed to restore partially~\cite{0407360}. This is a very important as well as a very hot topic. One of the main goals for RHIC (Relativistic Heavy Ion Collider) and LHC (Large Hadron Collider) is just to create and then study such a new state of matter. These studies also play crucial roles in researches on the evolution of the early universe. The properties of the QCD vacuum will change when the temperature and/or quark chemical potential vary, as a result the vacuum condensates and vacuum susceptibilities which reflect the nature of the QCD vacuum should also change with temperature and/or quark chemical potential. Theoretically, the QCD phase transitions via the calculations of temperature and quark chemical potential dependence of the two-quark condensate or QCD vacuum susceptibilities can then be studied. {In the standard definition, an order parameter is a quantity which is zero at one side of a phase transition and nonzero at the other side,
so that the two-quark condensate in the chiral limit is such a quantity for the chiral phase transition (therefore, in some cases it is also referred to as ``chiral condensate'')~\cite{PLB.420.267--273}. However, while away from the chiral limit the two-quark condensate is no longer an exact order parameter (since there is no exact phase transition), although it can be stretched to encompass first order transitions where the quantity has jumps from one nonzero value to another. Various susceptibilities, which are related to the derivatives of the two-quark condensate, are usually used to locate the transition, and hence are also important quantities to characterize the properties of the system.\footnote{Moreover, the two-quark condensate is actually unmeasurable, while various susceptibilities are reachable in the experiments.} For example, the scalar susceptibility is often used to describe the QCD chiral symmetry restoration at finite temperature and finite chemical potential~\cite{JHEP.09.073,PRD.77.076008,EPJC.56.483--492,PRD.71.034504,PRD.74.054507,PLB.643.46--54,Nature.443.675--678,PRD.75.034506,JHEP.06.088}, while studies of the temperature and chemical potential dependence of the quark number susceptibility can provide useful information to the vicinity of the critical end point (CEP), where the first order phase transition meets with the crossover and a second order phase transition takes place~\cite{PhysRevLett.59.2247,PRD.38.2888,PRD.40.2743,PLB.271.395,PRD.54.4585,PRD.55.6852,PRL.81.4816,PRD.64.074506,PRD.65.094515,PRD.65.054506,PRD.67.014028,
PhysRevD.67.034501,PRD.67.094018,PRD.68.014507,PRD.70.014016,PLB.649.57,PRD.76.076005,PRD.79.036001,PRC.79.035209,PLB.680.432--437,JHEP.02.066,PhysRevD.84.076004,JHEP.07.184}.} The cases with finite temperature and chemical potential represents a big branch in the field of studies on QCD vacuum susceptibilities. Hence it is difficult to give a comprehensive presentation on its theories and applications to the chiral phase transition of QCD. In this paper, we mainly discuss the calculations of various QCD vacuum susceptibilities at zero temperature and zero chemical potential, and then present some results in studies of the chiral phase transition of QCD. We will review the QCD sum rules external field formula in a nutshell, then take the constant vector field as an example to give the general derivation of the vector vacuum susceptibility, and obtain a model independent expression at last. We will also discuss the tensor, the vector, the axial-vector, the scalar, and the pseudo-scalar vacuum susceptibilities in detail together with their applications. From the model independent form of the vacuum susceptibilities, we see that their calculations are closely linked to the quark propagator (two-point Green function) and the corresponding vertex function (three-point Green function). {Lattice QCD, which is commonly accepted as a first principle calculation of QCD itself, can give good results for the two-point Green functions, but for three- and more points Green functions there is still a long way to go. Accordingly, in order to do the calculations we have to, at present, turn to some effective models or approaches that are based on QCD itself.} In this paper we mainly use the framework of DSEs approach, which is a very useful and successful non-perturbative method to treat the non-perturbative strong interactions as well as the properties of hadrons~\cite{PPNP.33.477,PLB.380.7--12,PRC.56.3369,PPNP.39.117--199,PRC.60.055214,PPNP.45.S1,PR.353.281--465,PRD.65.094026,IJMPE.12.297,PRD.70.014014,EPJA.18.231-235,Prl.96.022001,PPNP.61.50,PPNP.77.1-69}. It must be emphasized here that, DSEs is an infinite series of coupled non-perturbative integral equations that the Green functions of continuous space-time should satisfy, of which a given $n$-point Green function is related with the $m$-point Green function, where $m>n$. In order to solve practical problems, people have to take ``appropriate'' approximations to the DSEs.\footnote{A good approximation should holds the original symmetries of QCD itself as many as possible, and meanwhile, easy to calculate.} So far, the most commonly used approximation scheme is the rainbow-ladder approximation. The rainbow approximation means that in the DSE for quark propagator the full quark-gluon vertex is replaced by the bare vertex, while the ladder approximation is that in the Bethe-Salpeter equation (BSE) of the quark-meson vertex the quark--anti-quark scattering kernel is replaced by its lowest order perturbative contribution, details of which can be found in Refs.~\cite{PPNP.33.477,PRC.56.3369,PPNP.45.S1}. The rainbow-ladder approximation is self consistent~\cite{PLB.640.196--200}.\footnote{In the global color symmetry model (GCM)\cite{PPNP.39.117--199,PRD.32.2419}, if adopting the saddle point approximation, it can be found that the equation which quark propagator satisfy is just the unrenormalized DSE under rainbow approximation, while the one which quark-meson vertex satisfy is just the unrenormalized BSE under ladder approximation.} Besides, rainbow-ladder approximation also promises the DSEs to satisfy the axis vector WTI, which ensures that pion is the Goldstone boson, and its mass is zero in the chiral limit. Through many years of research, it has been found that the DSEs under rainbow-ladder approximation have achieved great success in describing the pseudo-scalar and the vector mesons, since in these cases the contributions from higher order correction terms is little~\cite{PRC.56.3369,PRC.60.055214}. However, the results for the scalar meson are not satisfactory, since in this case the higher order corrections contribute significantly ~\cite{PLB.380.7--12}. Therefore, for the scalar vertex and the scalar vacuum susceptibility the rainbow-ladder is not a good approximation. In other words, the results obtained in this approximation are not reliable, and one must go beyond it~\cite{PRC.79.035209}. Accordingly, in the following discussions the tensor vacuum susceptibility is within the rainbow-ladder approximation (the results for the BC vertex can be found in Ref.~\cite{FBS.48.31}, with a decrease of about 10\%), while for the scalar vacuum susceptibility we try to go beyond it.

\section[Derivation of the vector vacuum susceptibility]{Derivation of the vector vacuum susceptibility using the QCD sum rules external field formula}\label{sumrules}
In this section we derive a model independent expression of the vector vacuum susceptibility using the QCD sum rules external field formula. {To be specific, we will work in Euclidean space throughout this paper}. The vacuum susceptibilities are closely related to the linear responses of the full quark propagator to the external fields, so we add to the Lagrangian a term that the quark current coupled to an external field,
\begin{equation}
 J^{\Gamma}(y){\cal{V}}_{\Gamma}(y)\equiv\bar{q}(y)\Gamma q(y){\cal{V}}_{\Gamma}(y),
\end{equation}
here $J^{\Gamma}(y)$ is the quark current, ${\cal{V}}_{\Gamma}(y)$ is the external field, $q(y)$ is the quark field, $\Gamma$ is a matrix in the direct product space of Dirac, flavor, and color. The linear response term of the full quark propagator to the external field can be written as~\cite{PRC.72.035202},
\begin{equation}\label{eq:qk2Gjsumrule}
{\cal{G}}^{cc'\Gamma}_{\alpha\beta,q}(x)=\langle\tilde{0}|T[q^{c}_{\alpha}(x)\bar{q}^{c'}_{\beta}(0)]\tilde{0}\rangle_{J^{\Gamma}}={\cal{G}}^{cc'\Gamma,PT}_{\alpha\beta,q}(x)+{\cal{G}}^{cc'\Gamma,NP}_{\alpha\beta,q}(x),
\end{equation}
where $(cc')$ and $(\alpha\beta)$ are the color and spinor indices of the quark, $|\tilde{0}\rangle$ is the exact vacuum; {${\cal{G}}^{\Gamma,NP}_{q}(x)$ and ${\cal{G}}^{\Gamma,PT}_{q}(x)$ denote the non-perturbative and perturbative effects of the full quark propagator ${\cal{G}}^{cc'\Gamma}_{\alpha\beta,q}(x)$ under an external field, respectively. In the paper we adopt such a concept that,
${\cal{G}}^{cc'\Gamma}_{\alpha\beta,q}(x)$ is calculated in the Nambu vacuum, of which dynamical chiral symmetry breaking (DCSB) and quark confinement are two of the most important features; while ${\cal{G}}^{\Gamma,PT}_{q}(x)$ is calculated in the Wigner vacuum, where the system holds chiral symmetry (in the chiral limit case) and the dressed quarks are not confined (more discussions can be found in Sec.~\ref{qpqmv} of this paper, and Refs.~\cite{PRC.72.035202,PPNP.33.477,PPNP.45.S1}). As we will show in the following part, ${\cal{G}}^{\Gamma,NP}_{q}(x)$ is related to the vacuum susceptibilities and hard to calculate directly, while ${\cal{G}}^{cc'\Gamma}_{\alpha\beta,q}(x)$ and ${\cal{G}}^{\Gamma,PT}_{q}(x)$ can be calculated, and then we can get ${\cal{G}}^{\Gamma,NP}_{q}(x)$ as well as the vacuum susceptibilities indirectly.}

In the QCD sum rules theory, the vacuum susceptibility is defined as~\cite{PLB.129.328--334,NPB.232.109--142,PLB.271.223--230,PLB.576.289--296},
\begin{eqnarray}\label{eq:vacsuscQCDSumrule}
{\cal{G}}^{cc'\Gamma,NP}_{\alpha\beta,q}(x)&\equiv&\langle\tilde{0}|:q^{c}_{\alpha}(x)\bar{q}^{c'}_{\beta}(0):|\tilde{0}\rangle_{J^{\Gamma}} \nonumber\\
&\equiv&-\frac{1}{12}\left(\Gamma{\cal{V}}_{\Gamma}\right)_{\alpha\beta}\delta_{cc'}\chi^{\Gamma}H(x)\langle 0|:\bar{q}(0)q(0):|0\rangle ,
\end{eqnarray}
$|0\rangle$ represents the perturbative vacuum, $\chi^{\Gamma}$ is the vacuum susceptibility, while $H(x)$ is a function that characterize the non-local property of the non-local two-quark condensate, and $H(0)=1$. From Eq.~(\ref{eq:vacsuscQCDSumrule}) we can see that, in order to calculate $\chi^{\Gamma}$ we should know ${\cal{G}}^{cc'\Gamma,NP}_{\alpha\beta,q}(x)$ first, which means we should know ${\cal{G}}^{\Gamma,PT}_{q}(x)$ and ${\cal{G}}^{\Gamma}_{q}(x)$ simultaneously. Therefore, how to solve ${\cal{G}}^{\Gamma,PT}_{q}(x)$ and ${\cal{G}}^{\Gamma}_{q}(x)$ self-consistently becomes the key to get a model independent result. Next we will take a constant vector field ${\cal{V}}_{\mu}$ as an example to show how to get a model independent expression for the vector vacuum susceptibility~\cite{PRC.72.035202}.

When there is an external field, a term which is the coupling of the vector current and the vector external field should be added to the action of the system, namely,
$\Delta S\equiv\int d^4x~\bar{q}(x)i\gamma_{\mu}q(x){\cal{V}}_{\mu}(x)$. In this case, the dressed quark propagator in the chiral limit is
\begin{equation}
{\cal{G}}[{\cal{V}}](x)=\int{\cal{D}}\bar{q}{\cal{D}}q{\cal{D}}A~q(x)\bar{q}(0)\exp\left\{-S[\bar{q},q,A]-\int d^4x~\bar{q}(x)\gamma_{\mu}q(x){\cal{V}}_{\mu}(x)\right\},
\end{equation}
where
\begin{equation}
S[\bar{q},q,A]=\int d^4x\left\{\bar{q}({\not\!{\partial}}-ig\frac{\lambda^{a}}{2}{\not\!{A^a}})q+\frac{1}{4}F^{a}_{\mu\nu}F^{a}_{\mu\nu}\right\},
\end{equation}
and $F^{a}_{\mu\nu}=\partial_{\mu}A^{a}_{\nu}-\partial_{\nu}A^{a}_{\mu}+gf^{abc}A^{b}_{\mu}A^{c}_{\nu}$. Here we do not show the gauge fixing term, as well as the ghost field and its integration measure. Considering the linear response of the quark propagator to the external field, we can obtain
\begin{eqnarray}
{\cal{G}}[{\cal{V}}](x)&=&\int{\cal{D}}\bar{q}{\cal{D}}q{\cal{D}}A~ \left[q(x)\bar{q}(0)\right]~\exp\left\{-S[\bar{q},q,A]\right\}\nonumber\\
&&-\int{\cal{D}}\bar{q}{\cal{D}}q{\cal{D}}A~\left[q(x)\bar{q}(0)\int d^4y~ \bar{q}(y)\gamma_{\mu}q(y){\cal{V}}_{\mu}(x)\right]~ exp\left\{-S[\bar{q},q,A]\right\}+\cdot\cdot\cdot\nonumber\\
&\equiv&\langle\tilde{0}|T_{\tau}[q(x)\bar{q}(0)]|\tilde{0}\rangle-\int d^4y\langle\tilde{0}|T_{\tau}[q(x)\bar{q}(0)\bar{q}(y)\gamma_{\mu}q(y)]|\tilde{0}\rangle{\cal{V}}_{\mu}(y)+\cdot\cdot\cdot,
\end{eqnarray}
$T_{\tau}$ represents the time-ordering operation on Euclidean time $\tau$, $G(x)\equiv\langle\tilde{0}|T_{\tau}\left[q(x)\bar{q}(0)\right]|\tilde{0}\rangle={\cal{G}}[{\cal{V}}=0](x)$ is the full quark propagator without an external field (${\cal{V}}_{\mu}=0$), while
\begin{eqnarray}
{\cal{G}}^{{\cal{V}}}(x)\equiv\langle\tilde{0}|T_{\tau}[q(x)\bar{q}(0)]\tilde{0}\rangle_{{\cal{V}}}\equiv-\int d^4y\langle\tilde{0}|T_{\tau}[q(x)\bar{q}(0)\bar{q}(y)\gamma_{\mu}q(y)]|\tilde{0}\rangle {\cal{V}}_{\mu}(y)
\end{eqnarray}
is the linear response term of the quark propagator to the external field. Hence, now the main task is to know $\int d^4y\langle\tilde{0}|T_{\tau}[q(x)\bar{q}(0)\bar{q}(y)\gamma_{\mu}q(y)]|\tilde{0}\rangle$, we can try to do Taylor expansion of the inverse of dressed quark propagator to $\mathcal{V}_{\mu}$~\cite{PRC.72.035202,PRC.73.035206,PLB.639.248--257},
\begin{equation}
{\cal{G}}[{\cal{V}}]^{-1}=\left.{\cal{G}}[{\cal{V}}]^{-1}\right|_{{\cal{V}}_{\mu}=0} +\left.\frac{\delta{\cal{G}}^{-1}[{\cal{V}}]}{\delta {\cal{V}}_{\mu}}\right|_{{\cal{V}}_{\mu}=0}{\cal{V}}_{\mu}+\cdot\cdot\cdot\equiv G^{-1}+\Gamma_{\mu}{\cal{V}}_{\mu}+\cdot\cdot\cdot,
\end{equation}
and then only keep the first order term of the external field,
\begin{equation}\label{eq:GdV}
{\cal{G}}[{\cal{V}}]=G-G\Gamma_{\mu}{\cal{V}}_{\mu}G+\cdot\cdot\cdot.
\end{equation}
Here the vector vertex $\Gamma_{\mu}$ is defined as
\begin{equation}\label{eq:VGamma}
\Gamma_{\mu}(y_1,y_2;z)\equiv\left.\left[\frac{\delta{\cal{G}}[{\cal{V}}](y_1,y_2)^{-1}}{\delta{\cal{V}}_{\mu}(z)}\right]\right|_{{\cal{V}}_{\mu}=0}.
\end{equation}
The derivation of Eq.~(\ref{eq:GdV}) is general, so is applicable to both full and perturbative quark propagators. In configuration space it can be written explicitly as
\begin{eqnarray}
{\cal{G}}[{\cal{V}}](x)&=&G(x)-\int d^4 u_1 \int d^4 u_2~G(x,u_1)\Gamma_{\mu}(u_1,u_2)G(u_2,0){\cal{V}}_{\mu}+\cdot\cdot\cdot\nonumber\\
&=&G(x)-\int d^4 u_1 \int d^4 u_2~G(x,u_1)\left[\int\frac{d^4 p}{(2\pi)^4} e^{ip\cdot(u_1-u_2)}\Gamma_{\mu}(p,0)\right]G(u_2,0){\cal{V}}_{\mu}+\cdot\cdot\cdot\nonumber\\
&=&G(x)-\int\frac{d^4 p}{(2\pi)^4}e^{ip\cdot x}G(p)\Gamma_{\mu}(p,0)G(p){\cal{V}}_{\mu}+\cdots.
\end{eqnarray}
Then the linear response term of the full quark propagator ${\cal{G}}^{Z}(x)$ to an external field is,
\begin{eqnarray}\label{eq:Gexact1V}
{\cal{G}}^{{\cal{V}}}(x)\equiv\langle\tilde{0}|T_{\tau}\left[q(x)\bar{q}(0)\right]|\tilde{0}\rangle_{{\cal{V}}}=-\int\frac{d^4p}{(2\pi)^4}e^{ip\cdot x}G(p)\Gamma_{\mu}(p,0)G(p){\cal{V}}_{\mu}.
\end{eqnarray}
and similarly for the perturbative quark propagator ${\cal{G}}^{Z,PT}(x)$,
\begin{eqnarray}\label{eq:Gpt1V}
{\cal{G}}^{{\cal{V}},PT}(x)&\equiv&\langle 0|T_{\tau}\left[q(x)\bar{q}(0)\right]|0\rangle_{Z}\nonumber\\
&=&-\int d^4 u_1 \int d^4 u_2~G^{PT}(x,u_1)\left[\int\frac{d^4 p}{(2\pi)^4} e^{ip\cdot(u_1-u_2)}\Gamma^{PT}_{\mu}(p,0)\right]{\cal{V}}_{\mu}G^{PT}(u_2,0)\nonumber\\
&=&-\int\frac{d^4 p}{(2\pi)^4}e^{ip\cdot x}G^{PT}(p)\Gamma^{PT}_{\mu}(p,0){\cal{V}}_{\mu} G^{PT}(p),
\end{eqnarray}
where $G^{PT}(x)\equiv\langle 0|T_{\tau}\left[q(x)\bar{q}(0)\right]|0\rangle$, and $\Gamma^{PT}_{\mu}(p;0)$ is the corresponding perturbative vertex.

Now let's turn to the calculation of vacuum susceptibilities, the definition of the vector vacuum susceptibility $\chi^{{\cal{V}}}$ is,
\begin{equation}\label{eq:vvacsus00}
\langle\tilde{0}|:q^a_{\alpha}(0)\bar{q}^b_{\beta}(0):|\tilde{0}\rangle_{{\cal{V}}}^{NP}\cdot{\cal{V}}\equiv -\frac{(-i)}{12}(\gamma\cdot{\cal{V}})_{\alpha\beta}\delta_{ab}\chi^{{\cal{V}}}\langle\tilde{0}|:\bar{q}(0)q(0):|\tilde{0}\rangle,
\end{equation}
where ${\cal{G}}^{ab{\cal{V}},NP}_{\alpha\beta,q}(0)\equiv\langle\tilde{0}|:q^a_{\alpha}(0)\bar{q}^b_{\beta}(0):|\tilde{0}\rangle_{{\cal{V}}}^{NP}$ is the local two-quark condensate. It can be seen that the two-quark condensate is an explicitly gauge-invariant quantity, and is indeed computed in a gauge-invariant way for example in Lattice QCD simulations. That it depends on the renormalization scheme is on the other hand obvious.\footnote{{Without truncation, DSEs is equivalent to QCD; However, it is impossible to solve infinite coupled equations, to choose a better truncation scheme is then not only necessary but also challenging. Concerning the choice of gauge, we will use Landau gauge throughout this paper, which has many advantages~\cite{PPNP.33.477,PPNP.45.S1,PPNP.61.50,PPNP.77.1-69,AOP.267.1,AOP.324.106,CTP.58.79,PhysRevLett.109.252001,PhysRevC.85.045205}, for example, it is a fixed point of the renormalization group; that gauge for which sensitivity to model-dependent differences between Ansatze for the fermion-gauge boson vertex are least noticeable; and a covariant gauge, which is readily implemented in simulations of Lattice regularized QCD (see, e.g.,
Refs.~\cite{PhysRevD.71.054507,PLB.676.69,AIP.1354.45} and citations therein and thereto). Importantly, capitalisation on the gauge covariance of Schwinger functions obviates any question about the gauge dependence of gauge invariant quantities.}} If we multiply $\delta_{ab}\gamma_{\nu}^{\beta\alpha}$ on both sides of Eq.~(\ref{eq:vvacsus00}), then
\begin{eqnarray}\label{eq:vvacsus01}
&&\langle\tilde{0}|:\bar{q}(0)\gamma_{\nu}q(0):|\tilde{0}\rangle_{{\cal{V}}}^{NP}\cdot {\cal{V}}=(-i)\chi^{{\cal{V}}}{\cal{V}}_{\nu}\langle\tilde{0}|:\bar{q}(0)q(0):|\tilde{0}\rangle\nonumber\\
&&=\left[\langle\tilde{0}|T\left[\bar{q}(0)\gamma_{\nu}q(0)\right]|\tilde{0}\rangle_{{\cal{V}}}-\langle0|T\left[\bar{q}(0)\gamma_{\nu}q(0)\right]|0\rangle_{{\cal{V}}}^{PT}\right]\cdot{\cal{V}}.
\end{eqnarray}
From Eqs.~(\ref{eq:Gexact1V}) and (\ref{eq:Gpt1V}) we can find that
\begin{eqnarray}
\left[\langle\tilde{0}|T\left[\bar{q}(0)\gamma_{\nu}q(0)\right]|\tilde{0}\rangle_{{\cal{V}}}\right]\cdot{\cal{V}}&=&tr_{DC}\left[\gamma_{\nu}G\Gamma\cdot {\cal{V}} G\right],\label{eq:vectqqexact}\\
\left[\langle0|T\left[\bar{q}(0)\gamma_{\nu}q(0)\right]|0\rangle_{{\cal{V}}}^{PT}\right]\cdot{\cal{V}}&=&tr_{DC}\left[\gamma_{\nu}G^{PT}\Gamma^{PT}\cdot {\cal{V}}G^{PT}\right].\label{eq:vectqqpt}
\end{eqnarray}
substituting them into Eq.~(\ref{eq:vvacsus01}), we can get a QCD sum rules based, also model independent expression of the vector vacuum susceptibility as,
\begin{equation}
\chi^{{\cal{V}}}\langle\tilde{0}|:\bar{q}(0)q(0):|\tilde{0}\rangle=i\left[tr_{DC}[\gamma_{\nu}G\Gamma_{\mu}G] -tr_{DC}[\gamma_{\nu}G^{PT}\Gamma^{PT}_{\mu}G^{PT}]\right]\frac{{\cal {V}_{\mu}}{\cal {V}_{\nu}}}{{\cal{V}}^2}.\label{sus-ge}
\end{equation}
Here we note that, from the viewpoint of Feynman diagram, $tr_{DC}[\gamma_{\nu}G\Gamma_{\mu} G]$ and $tr_{DC}[\gamma_{\nu}G^{PT}\Gamma^{PT}_{\mu} G^{PT}]$ are respectively the values of the full and the perturbative vector vacuum polarizations at zero momentum, so the calculation of the vector vacuum susceptibility comes down to the calculation of the vector vacuum polarization. Here we want to stress that, as pointed out in Ref.~\cite{PRC.72.035202}, the derivation in this section is general and model independent for a constant external field, and so it can also be used to derive other vacuum susceptibilities similarly, such as the tensor vacuum susceptibility. However, for the axial-vector vacuum susceptibility, we need to take a variable external field, details can be found in Ref.~\cite{PRC.73.035206}. The significance of the works in Refs.~\cite{PRC.72.035202,PRC.73.035206} is that they give theoretically model independent results of various vacuum susceptibilities, which provides a good starting point for further model calculations. In the next section we will discuss the tensor, the vector, the axial-vector, the scalar, and the pseudo-scalar vacuum susceptibilities respectively.

\section{Calculations of various vacuum susceptibilities}\label{calsus}
After the model independent expressions of QCD vacuum susceptibilities are obtained, the next step is to use these expressions to calculate the values of these vacuum susceptibilities. From Eq.~(\ref{sus-ge}) we know that, in order to calculate the vector vacuum susceptibility, we must first know the quark propagator $G(P)$ (two-point Green function) and the vertex function $\Gamma_{\mu}(P, 0) $(three-point Green function). This holds for the other vacuum susceptibilities. However, as we mentioned earlier, {nowadays} it is difficult or even impossible to know these Green functions from the first principles of QCD directly, and people have to resort to various non-perturbative approaches {besides the Lattice QCD}, such as the QCD sum rules, the chiral perturbation theory, the DSEs theory, etc. Generally speaking, each approach has its own advantages and disadvantages. Since the early ninety's of the last century, the DSEs has made great progress, and also gained a lot of results which are in good agreement with the experimental measurements and Lattice QCD calculations~\cite{PPNP.33.477,PPNP.45.S1,PR.353.281--465,PRD.65.094026,IJMPE.12.297,PRD.70.014014,EPJA.18.231-235,Prl.96.022001,PPNP.61.50}. This approach, therefore, is getting more and more attentions in many fields. In the following part of this section, we will discuss the calculations of the quark propagator and the quark-meson vertex in the framework of DSEs, and then discuss the calculations of several vacuum susceptibilities.

\subsection[Calculations of quark propagator and quark-meson vertex]{Calculations of quark propagator and quark-meson vertex in the framework of the DSEs}\label{qpqmv}
Using the functional path integral techniques, we can get the DSEs for Green functions from the Lagrangian (details can be found in the textbooks of Quantum Field Theory). Since the derivation of functional path integral does not rely on the perturbative theory, DSEs is strict (in the weak coupling limit, DSEs give the standard Feynman diagram expansion of the Green functions). The DSE of the renormalized quark propagator is
\begin{eqnarray}
G(p)^{-1}=Z_2(i\gamma\cdot p+m_{bm})+Z_1\int^{\Lambda}_{q}g^2D_{\mu\nu}(p-q)\gamma_{\mu}\frac{\lambda^{a}}{2}G(q)\Gamma^a_{\nu}(q,p),\label{quark-pro}
\end{eqnarray}
here $g^2_{s}D^{ab}_{\mu\nu}(p)$ is the renormalized gluon propagator, {$\lambda^{a}$ is the Gell-Mann matrices with $a$ the color index}, $\Gamma^a_{\nu}(q,p)$ is the renormalized quark-gluon vertex, $m_{bm}$ is the bare mass of current quark and is related to $\Lambda$, $\int^{\Lambda}_{q}\equiv\int^{\Lambda}d^4q/(2\pi)^4$ represents some regularization scheme that keeps translation invariance, where $\Lambda$ is the mass scale, we should take the limit $\Lambda\rightarrow\infty$ at last. $Z_1(\mu^2,\Lambda^2)$ and $Z_2(\mu^2,\Lambda^2)$ are the renormalization constants of the quark-gluon vertex and quark wave function, which depend on both the renormalization point $\mu$ and the renormalization mass scale $\Lambda$. According to the Lorentz structure analysis, the solution of the quark DSE Eq.~(\ref{quark-pro}) has the following form
\begin{equation}
G(p)^{-1}\equiv i\gamma\cdot pA(p^2,\mu^2)+B(p^2,\mu^2).\label{quark-gf}
\end{equation}
Eq.(\ref{quark-pro}) must be solved under certain renormalization condition. Since the asymptotic freedom nature of QCD, people usually demand that at a big space-like momentum square there should be
\begin{equation}
\left.G(p)^{-1}\right|_{p^2=\mu^2}=i\gamma\cdot p+m(\mu),\label{renor-cond}
\end{equation}
where $m(\mu)$ is the renormalized current quark mass at $\mu$. When chiral symmetry is explicitly broken, the renormalized mass and the bare mass satisfy this relation, $m(\mu)=m_{bm}(\Lambda)/Z_m(\mu^2,\Lambda^2)$, here $Z_m(\mu^2,\Lambda^2)$ is the renormalization constant of mass; while there is no explicit chiral symmetry breaking, $Z_2m_{bm}=0$, which is just the case of chiral limit. Most of the following discussions on the vacuum  susceptibilities are for the chiral limit case.

It can be seen clearly from Eq.~(\ref{quark-pro}) that, the quark propagator, the gluon propagator, and the quark-gluon vertex are related to each other. The gluon propagator and the quark-gluon vertex also satisfy their own DSEs, which are coupled to the Green functions with more points. Therefore, in order to obtain a closed equation of the quark propagator, we must make some appropriate approximations to the gluon propagator and the quark-gluon vertex. On the gluon propagator side, the researches of its DSE have made great progress, {and the gluon propagator in Landau gauge is for all phenomenological purposes sufficiently well-known~\cite{PPNP.33.477,PPNP.45.S1,PPNP.61.50,PPNP.77.1-69,CTP.58.79,AOP.267.1,PhysRevLett.109.252001,AOP.324.106,PhysRevC.84.042202,PhysRevD.86.014032,PLB.742.183}. The lack of knowledge refers much more to the quark-gluon vertex. It is generally believed that, in studies of the color confinement, the DCSB, and the properties of bound states, the non-perturbative quark-gluon vertex plays an important role~\cite{JHEP.09.013,JHEP.04.047,PRC.70.035205,PRD.70.094039,NPBS.141.244--249,PRD.73.094511,EPJC.50.871--875}. One of the most complete investigations of the quark-gluon vertex can be found in Ref.~\cite{AOP.324.106}, and further guides as well as some important recent works and studies beyond the rainbow-ladder truncation can be found in Refs.~\cite{JHEP.10.193,PhysRevD.90.065027,1404.2545} (and references therein)}. Therefore, when studying the dynamical chiral symmetry breaking problems of the quark propagator, people usually choose an appropriate gluon propagator model as input. A gluon propagator model often contains some parameters, which can be fixed by fitting the low-energy hadron physics experiments (such as the mass and decay constant of pion). For the quark-gluon vertex, the case is more complicated. The most simple approximation is just the rainbow approximation, namely, use the bare quark-gluon vertex $\gamma_{\mu}$ to replace the full one $\Gamma_{\mu}(q, p)$. This is a big approximation, since the momentum dependence of the quark-gluon vertex is completely neglected, and of course, the non-perturbative information is also lost completely. In studies of low-energy hadron physics, such as the calculation of hadron masses, decay constants, and other physical quantities, it is necessary to solve the BSE that the quark-meson vertex satisfy. As mentioned above, the rainbow approximation of the quark DSE and the ladder approximation of the quark-meson vertex BSE are consistent with each other, and over many years of studies it has been demonstrated that although the rainbow-ladder approximation is the most simple one, it can give successful descriptions to some meson properties. The reason why it can not describe the scalar meson well has also become known. There have been many attempts that try to study the scalar meson beyond the rainbow-ladder approximation. However, this is rather difficult work, since if we do some approximations to the quark-gluon vertex of the quark DSE,\footnote{For example, by the constraint of the longitudinal WTI and the requirement that the vertex has no kinematic singularity, people have developed the Ball-Chiu (BC) and Curtis-Pennington (CP) approximations for fermion-boson vertex in the gauge theory~\cite{PRD.22.2542,PRD.22.2550,PRD.42.4165}. In recent years, people also try to use the constraint of the transverse WTI~\cite{PLB.480.222--228,PRC.63.025207}.} we also need to find a corresponding quark-meson BSE that is consistent with these approximations, which is really a difficult work. As we mentioned earlier, for studies of scalar vacuum susceptibility people should go beyond the rainbow-ladder approximation. The authors of Ref.~\cite{PRC.79.035209} noted that in the calculations of the scalar vacuum susceptibility, only the scalar vertex with zero momentum exchange $\Gamma(p, 0)$ is used, rather than the full scalar vertex $\Gamma(p, q)$. Using this feature and the WTI, they give a method that can describe the scalar vacuum susceptibility beyond the rainbow-ladder approximation. In the following we will introduce the equations of the quark propagator and the quark-meson vertex under rainbow-ladder approximation first, then present a method to go beyond it.

Under the rainbow approximation and taking the chiral limit ($m_{bm}=0$), and substituting Eq.~(\ref{quark-gf}) into Eq.~(\ref{quark-pro}), it is easy to know the equations that $A (p^2, \mu^2)$ and $B (p^2, \mu^2)$ satisfy
\[
[A(p^2,\mu^2)-Z_2]p^2=\frac{4}{3}\int^{\Lambda}_{q}\frac{g^2_{s}D(p-q)A(q^2,\mu^2)}{q^2A^2(q^2,\mu^2)+B^2(q^2,\mu^2)}\left[p\cdot q+2\frac{p\cdot(p-q)~q\cdot(p-q)}{(p-q)^2}\right],
\]
\begin{equation}
B(p^2,\mu^2)=4\int^{\Lambda}_{q}\frac{g^2_{s}D(p-q)B(q^2,\mu^2)}{q^2A^2(q^2,\mu^2)+B^2(q^2,\mu^2)},\label{Bfunc}
\end{equation}
where we take the Landau gauge for the gluon propagator $D^{ab}_{\mu\nu}(p)=\delta^{ab}\left(\delta_{\mu\nu}-p_{\mu}p_{\nu}/{p^2}\right)D(p^2)$. After inputting the model gluon propagator, Eq. (\ref{Bfunc}) can be solved by numerical iteration. From Eq.~(\ref{Bfunc}) we can see clearly that it has two distinct solutions: the Nambu-Goldstone solution (or Nambu solution) that $B(p^2, \mu^2)\neq0$ and the Wigner-Weyl solution (or Wigner solution) that $B(p^2, \mu^2)\equiv0$. In the Nambu-Goldstone phase: 1) chiral symmetry breaking occurs dynamically, the originally massless current quarks obtain masses by DCSB; 2) quarks are confined, since the quark propagator does not have the Lehmann representation. While in the Wigner phase, there is no DCSB, and the quarks are not confined~\cite{PPNP.33.477,PPNP.39.117--199,PPNP.45.S1}.

In the Wigner phase, Eq.~(\ref{Bfunc}) becomes,
\begin{equation}
[A'(p^2,\mu^2)-Z_2']p^2=\frac{4}{3}\int^{\Lambda}_{q}\frac{g_{s}^2D(p-q)}{q^2A'(q^2,\mu^2)}\left[p\cdot q+2\frac{p\cdot(p-q)~q\cdot(p-q)}{(p-q)^2}\right],
\end{equation}
here $A'(p^2,\mu^2)$ represents the vector part of the self-energy function. Therefore, the chiral limit case of the quark propagator in the Wigner phase is
\begin{equation}
G^{(W)}(p)=\frac{-i\gamma\cdot p}{A'(p^2,\mu^2)p^2}.
\end{equation}

In the ladder approximation, quark-meson vertex (three-point Green function) also satisfies its own BSE. For the tensor vertex, its BSE is
\begin{equation}
\Gamma^m_{\mu\nu}(p,0)=Z_T\sigma_{\mu\nu}-\frac{4}{3}\int^{\Lambda}_{q}g^2_{s}D_{\eta\zeta}(p-q)\gamma_{\eta}G(q)\Gamma^m_{\mu\nu}(q,0)G(q)\gamma_{\zeta},\label{vertex-ten}
\end{equation}
{where the superscript $m$ denote ``meson'', and then $\Gamma$ means quark-gluon vertex while $\Gamma^m$ represents quark-meson vertex.} Similar to the quark propagator, we choose such a the renormalization condition
\begin{equation}
\left.\Gamma^m_{\mu\nu}(p,0)\right|_{p^2=\mu^2}=\sigma_{\mu\nu}.\label{renor-cond-vertex}
\end{equation}
According to Lorentz structure analysis, the general form of $\Gamma^m_{\mu\nu}$ can be written as the following
\begin{eqnarray}
\Gamma^m_{\mu\nu}(p,0)&=&\sigma_{\mu\nu}\Lambda_1(p^2,\mu^2)+\sigma_{\mu\gamma}p_{\gamma}p_{\nu}\Lambda_2(p^2,\mu^2)+\gamma_{\mu}p_{\nu}\Lambda_3(p^2,\mu^2)\nonumber\\
&&+\gamma\cdot pp_{\mu}p_{\nu}\Lambda_4(p^2,\mu^2)+ip_{\mu}p_{\nu}\Lambda_5(p^2,\mu^2).\label{vertex-lorentz}
\end{eqnarray}
Now substituting Eq.~(\ref{vertex-lorentz}) into Eq.~(\ref{vertex-ten}) and after some algebra, we get five coupled integral equations of the scalar function $\Lambda_i(p^2,\mu^2)(i=1,\cdot\cdot\cdot,5)$, which have the solution $\Lambda_{3,4,5}(p^2,\mu^2)\equiv0$. So that~\cite{PLB.639.248--257}
\begin{equation}
\Gamma^m_{\mu\nu}(p,0)=\sigma_{\mu\nu}\Lambda_1(p^2,\mu^2)+\sigma_{\mu\gamma}p_{\gamma}p_{\nu}\Lambda_2(p^2,\mu^2),
\end{equation}

The BSE that the vector vertex satisfy is
\begin{equation}
\Gamma^m_{\mu}(p,0)=Z_2\gamma_{\mu}-\frac{4}{3}\int^{\Lambda}_{q}g^2_{s}D_{\eta\zeta}(p-q)\gamma_{\eta}G(q)\Gamma^m_{\mu}(q,0)G(q)\gamma_{\zeta}.\label{vertex-vect}
\end{equation}
Here one way to solve this is to write down the general Lorentz structure of $\Gamma^m_{\mu}(p,0)$ and then do the numerical iteration, another more direct way is to use the Ward identity ({which is obtained via the Ward-Takahashi identity by taking the infrared limit of the photon})
\begin{equation}
\Gamma^m_{\mu}(p,0)=-\frac{\partial  G(p)^{-1}}{\partial p_{\mu}},
\end{equation}
by calculating the partial derivative of the quark propagator to the momentum $p$, we can know the vector vertex directly. Similar treatment can also be used to the axial-vector vertex, for the details please see Ref.~\cite{PRC.73.035206}.

For the scalar vertex, the rainbow-ladder approximation is not a good one and we have to go beyond it. Let us see its BSE first,
\begin{equation}
\Gamma^m(p,0)=1+\int\frac{d^4q}{(2\pi)^4}K(q,p)[G(q)\Gamma^m(q,0)G(q)],\label{scalarbs}
\end{equation}
here $K(q,p)$ is the quark--anti-quark scattering kernel, which is a four-point Green function that contains many non-perturbative information. As mentioned above, when beyond the rainbow-ladder approximation it is very difficult to find the corresponding BSE of the vertex. But if take a closer look we can find that only the scalar vertex ({it is known from the functional path integral theory that taking the derivative of the two-point Green function with respect to the scalar external field will give the scalar vertex}) $\Gamma^m(p,0)=\frac{\partial G(p)^{-1}}{\partial m}|_{m=0}$ is needed to calculate the scalar vacuum susceptibility. The authors of Ref.~\cite{PRC.79.035209} skillfully used this point. By the DSE of quark propagator, Eq.~(\ref{quark-pro}), it is easy to know the equation of scalar vertex
\[
\Gamma^m(p,0)=1+\Sigma_1(p)+\Sigma_2(p),
\]
here
\begin{eqnarray}
\Sigma_1(p)&=&-\frac{4}{3}\int_{q}^{\Lambda}g^2D_{\mu\nu}(p-q)\gamma_{\mu}G(q)\Gamma^m(q,0)G(q)\Gamma_{\nu}(q,p),\nonumber\\
\Sigma_2(p)&=&\frac{4}{3}\int_{q}^{\Lambda}g^2D_{\mu\nu}(p-q)\gamma_{\mu}G(q)\Lambda_{\nu}(q,p),\label{vertex}
\end{eqnarray}
{where the summation over color is carried out to be $4/3$} and $\Lambda_{\nu}(q,p)=\frac{\partial\Gamma_{\nu}(q,p)}{\partial m}$.

Using the vector Ward-Takahashi identity,\footnote{{Actually, the QCD vertex does not satisfy this identity, but rather the more complicated Slavnov-Taylor identity (STI), which can be derived from the Becchi-Rouet-Stora invariance of QCD and corresponds to the WTI of QED. The difficult challenge is to find an Ansatz for the unknown renormalised propagators, vertices, etc., of QCD which satisfy the DSEs and respect the STIs of the theory. Progress can be made by satisfying a subset of the DSEs and STIs, and then supplementing these with information gleaned from Lattice QCD, phenomenology, etc. However, without ghosts Eq.~(\ref{wti}) is identical to the corresponding WTI of QED~\cite{PPNP.33.477}.}}
\begin{equation}
ik_\nu\Gamma_{\nu}(q,p)=G(q)^{-1}-G(p)^{-1},~~~~~ k=q-p\label{wti}
\end{equation}
we can know another identity that $\Lambda_{\nu}(q,p)$ and $\Gamma^m(p)$ satisfy,
\begin{equation}
ik_\nu\Lambda_{\nu}(q,p)=\Gamma^m(q,0)-\Gamma^m(p,0).\label{vwti}
\end{equation}
This is an important step towards the consistent solutions of quark propagator and the scalar vertex.

By the Lorentz structure analysis, the inverse of the quark propagator $G(p)^{-1}$ and the scalar vertex $\Gamma$ has such general forms,
\begin{equation}
G(p)^{-1}=i\gamma\cdot pA(p^2)+B(p^2),\label{lorentzq}
\end{equation}
\begin{equation}
\Gamma^m(p,0)=i\gamma\cdot pC(p^2)+D(p^2).\label{lorentzv}
\end{equation}
And according to the works of Ball and Chiu~\cite{PRD.22.2542,PRD.22.2550}, we take the following forms of $\Gamma_{\nu}(q,p)$ and $\Lambda_{\nu}(q,p)$
\begin{eqnarray}
\label{bcvertex}
\Gamma_{\mu}(q,p)&=&\Sigma_A\gamma_\mu+(q+p)_\mu[\frac{1}{2}\gamma\cdot(q+p)\Delta_A-i\Delta_B]\nonumber\\
\Lambda_{\mu}(q,p)&=&\Sigma_C\gamma_\mu+(q+p)_\mu[\frac{1}{2}\gamma\cdot(q+p)\Delta_C-i\Delta_D]\nonumber,
\end{eqnarray}
where
\begin{equation}
\Sigma_{\cal F}=\frac{1}{2}[{\cal F}(q^2)+{\cal F}(p^2)],~~\Delta_{\cal F}=\frac{{\cal F}(q^2)-{\cal F}(p^2)}{q^2-p^2},\label{bcvertex}
\end{equation}
and ${\cal F}=A,B,C,D$. Here it should be pointed out that, there are also many other forms of vertex ansatz that can also satisfy the vector WTI. The main reasons that the authors of Ref.~\cite{PRC.79.035209} chose the BC vertex is that: firstly, there is no kinematic singularity; secondly, the momenta $p$ and $q$ are symmetrical. The main purpose of their work is to discuss the differences of the results between the bare vertex and a corrected one.

Now substituting $\Gamma_{\nu}(q, p) $ into the DSE of the quark propagator Eq.~(\ref{quark-pro}), we can find the numerical solutions of the vector function $A$ and the scalar function $B$ of the quark propagator. Then by substituting $\Gamma_{\nu}(q, p)$, $\Lambda_\nu(q, p)$, $A$, and $B$ into Eq.~(\ref{vertex}) we can solve the scalar vertex. This method can be generalized to the calculations of vacuum susceptibilities at finite temperature and finite quark chemical potential. In the following part we will show the details of the gluon propagators and the numerical results.

\subsection{Various vacuum susceptibilities}
In this part we discuss the calculations and results of various vacuum susceptibilities in detail, namely, the tensor, the vector, the axial-vector, the scalar, and the pseudo-scalar vacuum susceptibilities.

\subsubsection{The tensor vacuum susceptibility}
Studies of tensor vacuum susceptibility are closely related to the tensor charge of nucleon~\cite{NPB.375.527--560}. {There are some calculations related to the tensor vacuum susceptibility, however, the results have shown that the theoretical treatments of this quantity is subtle and different approaches may lead to different outputs, even have different sign. For example, within the framework of QCD sum rules, H. X. He and X. D.Ji (1995, 1996) got about 0.002~GeV$^2$~\cite{PHYSREVD.52.2960,PRD.54.6897}, V. M. Belyaev and A. Oganesian (1997) got -0.008~GeV$^2$~\cite{PLB.395.307--310}, L. S. Kisslinger (1999) got 0.0072 $\sim$ 0.0127~GeV$^2$~\cite{PRC.59.3377}, A. P. Bakuleva and S. V. Mikhailov (2000) got $-0.0055\pm0.0008$~GeV$^2$ for the non-local condensates sum rule and $-0.0053\pm0.0021$~GeV$^2$ for standard sum rule~\cite{EPJC.17.129--135}; while W. Broniowski $et~al$ (1998) got -(0.0083 $\sim$ 0.0104)~GeV$^2$ using the chiral constituent model~\cite{PLB.438.242--247}, and H. T. Yang $et~al$ (2003) got -(0.0014 $\sim$ 0.0020)~GeV$^2$ within global color symmetry model~\cite{PLB.557.33--37}.} Clearly, in order to get a reliable theoretical prediction of the tensor charge, we should determine the tensor vacuum susceptibility as precise as possible. The work of Ref.~\cite{PLB.639.248--257} has provided a general treatment of the issue, and in Ref.~\cite{FBS.48.31} the calculations have been performed by employing the Ball-Chiu (BC) vertex~\cite{PRD.22.2542,PRD.22.2550}. The results of Ref.~\cite{FBS.48.31} show that the tensor vacuum susceptibility obtained in the BC vertex approximation is reduced by about 10\% compared to its rainbow-ladder approximation value, which means that the dressing effect of the quark-gluon vertex is not very large in the calculation of the tensor vacuum susceptibility within the DSEs framework.

Through the previous general discussions of the vacuum susceptibilities (Sec.~\ref{sumrules} of this paper) we know that, if we introduce a term that the quark tensor current coupled to a constant tensor external field ${\bar q}(x)\sigma_{\mu\nu}q(x)Z_{\mu\nu} (Z_{\mu\nu}$ is the tensor external field), using the QCD sum rules approach we can obtain the general expression of the tensor vacuum susceptibility (here is the chiral limit case, but it is easy to generalize to the case with nonzero current quark mass)
\begin{eqnarray}\label{tensor1}
\chi^{Z}&=&\frac{\left\{Tr[\sigma_{\eta\zeta}G\Gamma\cdot Z G]-Tr[\sigma_{\eta\zeta}G^{PT}\Gamma^{PT}\cdot Z G^{PT}]\right\}}{Z_{\eta\zeta}\langle\tilde{0}|:\bar{q}(0)q(0):|\tilde{0}\rangle}.
\end{eqnarray}
The expression in the earlier literatures is~\cite{PLB.438.242--247,PRC.59.3377,EPJC.17.129--135,PLB.557.33--37}
\begin{eqnarray}\label{tensor2}
\chi'^{Z}&=&\frac{\left\{Tr[\sigma_{\eta\zeta}G\sigma\cdot Z G]-Tr[\sigma_{\eta\zeta}G^{PT}\sigma\cdot Z G^{PT}]\right\}}{Z_{\eta\zeta}\langle\tilde{0}|:\bar{q}(0)q(0):|\tilde{0}\rangle},
\end{eqnarray}
Comparing these two equations it is easy to see that, the difference between $\chi^{Z}$ and $\chi'^{Z}$ comes from the fact that in the previous calculations instead of using the exact dressed vertex $\Gamma_{\mu\nu}$ and the perturbative one $\Gamma^{PT}_{\mu\nu}$, they used the bare vertex $\sigma_{\mu\nu}$, in the course of which the non-perturbative and perturbative dressing effects on the vertex were ignored. Here we want to stress that, both the full and the perturbative tensor vertexes have complicated momentum dependence. Especially for the full tensor vertex, there may exist mass singularities of various meson bound states, so compared with the bare vertex it contains quite a lot of sophisticated non-perturbative effects of QCD.

The discussions in Ref.~\cite{PLB.639.248--257} are under the rainbow-ladder approximation, and the renormalization procedure is necessary. The renormalized version of the tensor vacuum susceptibility is (for the details please see Ref.~\cite{PLB.639.248--257})
\begin{equation}\label{renor-tensor}
\chi^{Z}_R=\frac{\left\{Tr[\sigma_{\eta\zeta}G_R\Gamma_R\cdot Z G_R]-Tr[\sigma_{\eta\zeta}G^{PT}_R\Gamma^{PT}_R\cdot Z G^{PT}_R]\right\}} {Z_T Z_{\eta\zeta}\langle\tilde{0}|:\bar{q}_R(0)q_R(0):|\tilde{0}\rangle},
\end{equation}
where $G_R$, $\Gamma_R$, $G^{PT}_R$ and $\Gamma^{PT}_R$ represent the renormalized versions of the full quark propagator, the full tensor vertex, the perturbative quark propagator and the perturbative tensor vertex, respectively. $\langle\tilde{0}|:\bar{q}_R(0)q_R(0):|\tilde{0}\rangle$ is the renormalized two-quark condensate.

To show the numerical difference between $\chi^{Z}$ and $\chi'^{Z}$, we need to specify a model gluon propagator. The one Ref.~\cite{PLB.639.248--257} selected is the popular as well as famous Maris-Tandy model~\cite{PRC.60.055214}, which reads
\begin{equation}
g^2_sD(k^2)=\frac{4\pi^2}{\omega^6}Dk^2e^{-k^2/\omega^2}+4\pi\frac{\gamma_m\pi}{\frac{1}{2}\ln\left[\tau+\left(1+k^2/\Lambda^2_{QCD}\right)^2\right]}{\cal{F}}(k^2),
\end{equation}
here ${\cal{F}}=\left[1-\exp(-k^2/[4m^2_t])\right]/k^2$, $\tau=e^2-1$, $\gamma_m=12/(33-2N_f)$ is the anomalous dimension of mass. For $N_f=4$, $\Lambda^{N_f=4}_{QCD}=0.234~\mathrm{GeV}$, and the renormalization point is chosen as $\mu=19~\mathrm{GeV}$. As Ref.~\cite{PRC.60.055214} pointed out, the parameters $\omega$ and $m_t$ are not independent, to chose $m_t=0.5~\mathrm{GeV}$ and $\omega=0.3~\mathrm{GeV}$ is to ensure that in the region $2~\mathrm{GeV}^2\le k^2$ there is $g_s^2D(k^2)\simeq 4\pi\alpha_s(k^2)$, where $\alpha_s(k^2)$ is the ``running'' coupling constant of QCD. In Ref.~\cite{PRC.60.055214} they chose $D=1.25~\mathrm{GeV}^2$, and there studies show that this set of parameters can give very good results for the properties of pion and kaon.

\begin{figure}[htb]
\centering
\includegraphics[width=0.7\textwidth]{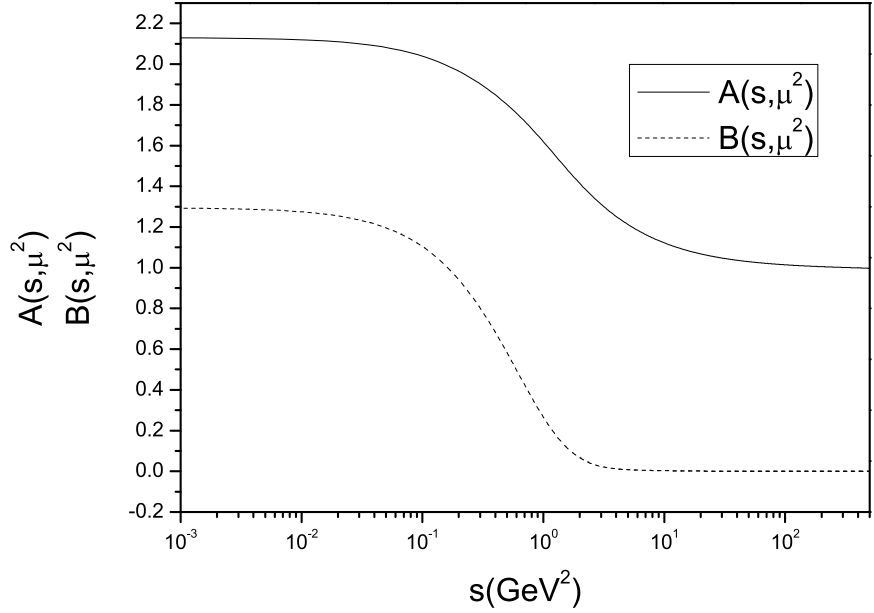}
\caption{The scalar functions $A(s,\mu^2)$ and $B(s,\mu^2)$ {[GeV]} of the dressed quark propagator, taken from Ref.~\cite{PLB.639.248--257}. {Here $s=p^2$}.}\label{nab}
\end{figure}

\begin{figure}[htb]
\centering
\includegraphics[width=0.49\textwidth]{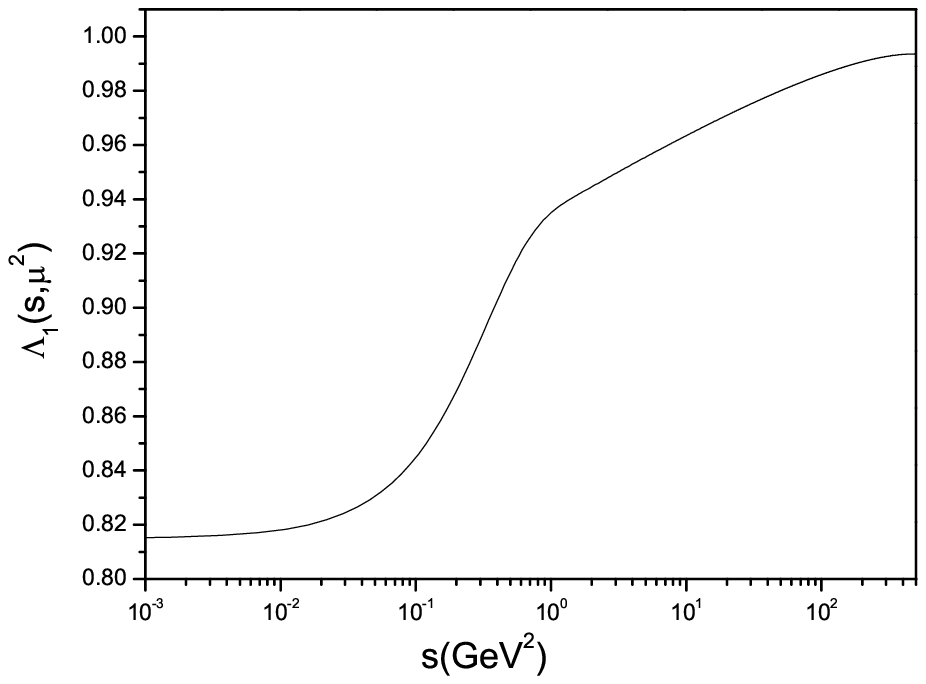}
\includegraphics[width=0.49\textwidth]{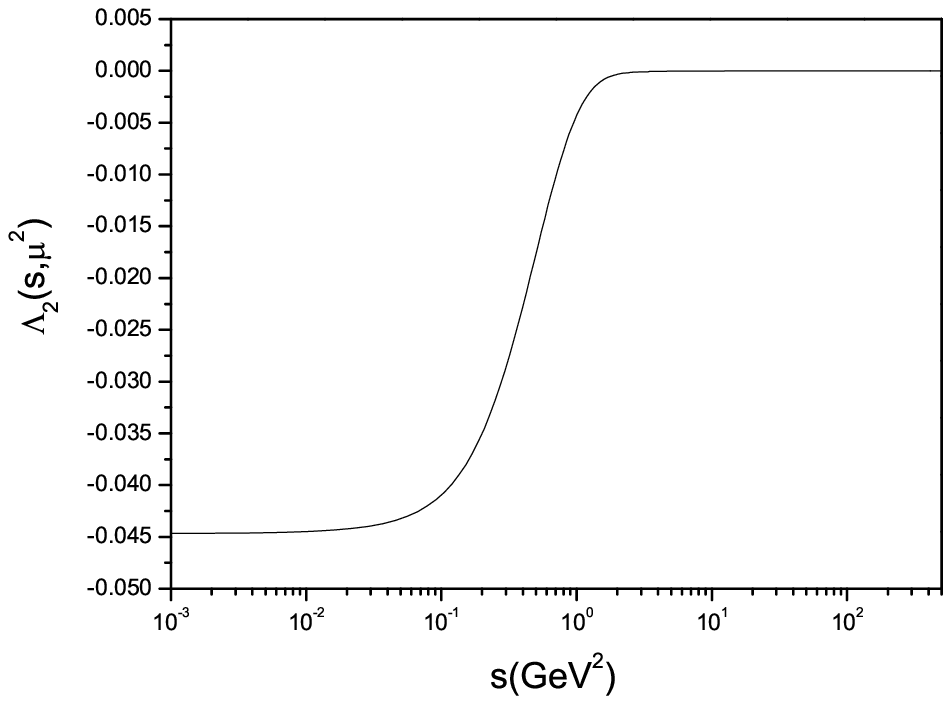}
\caption{\label{nLamda} The tensor vertex functions $\Lambda_1(s,\mu^2)$ and $\Lambda_2(s,\mu^2) ${[GeV$^{-2}$]}, taken from Ref.~\cite{PLB.639.248--257}. {Here $s=p^2$}.}
\end{figure}

Once the model gluon propagator is chosen, we can then calculate the quark propagator and the vertex through numerical iteration, as shown in Fig.~\ref{nab} and Fig.~\ref{nLamda} respectively. The final numerical results of the tensor vacuum susceptibilities are
\begin{equation}
\chi^{Z} a=-0.0027~\mathrm{GeV}^2~~, ~~\chi'^{Z}a=-0.0031~\mathrm{GeV}^2,
\end{equation}
here $a\equiv-\frac{1}{2}\langle\tilde{0}|:\bar{q}_R(0)q_R(0):|\tilde{0}\rangle$.

We see that, $\chi^{Z}$ and $\chi'^{Z}$ are indeed different, but the difference is not very large. Let us now analyze the reasons for this. As stated above, the difference between them comes from the choices of vertexes. We know that in the perturbative region the bare vertex $\sigma_{\mu\nu}$ can be regarded as the zeroth approximation of the dressed vertex $\Gamma_{\mu\nu}$, and therefore it is physically interesting to analyze the effects caused by their differences in the non-perturbative region. It can be seen clearly from Fig.~\ref{nLamda} that, for large enough momentum square $s=p^2$, $\Lambda_1(s,\mu^2)$ tends to 1 while $\Lambda_2(s,\mu^2)$ tends to 0, which means that in the large momentum region $\Gamma_{\mu\nu}$ tends to $\sigma_{\mu\nu}$; however, for small $s$, $\Lambda_1(s,\mu^2)$ differs from 1 and $\Lambda_2(s,\mu^2)$ differs from 0, although the differences are not significant. Therefore, the dressed vertex $\Gamma_{\mu\nu}(p,0)$ obtained in the rainbow-ladder approximation of the DSEs approach is similar to that of the bare vertex in the small momentum region, and it is just this small difference that accounts for the small numerical difference between the numerical values of $\chi^{Z}$ and $\chi'^{Z}$. The small magnitude of the numerical difference between $\chi^{Z}$ and $\chi'^{Z}$ suggests that the non-perturbative dressing effects on the dressed tensor vertex are not important, and therefore in this case to use $\sigma_{\mu\nu}$ as a replacement of $\Gamma_{\mu\nu}(p,0)$ is a rather good approximation. This is quite different from the case of vector vacuum susceptibility~\cite{PRC.72.035202}.

\subsubsection{The vector and axial-vector vacuum susceptibilities}
For the vector vacuum susceptibility $\chi_{V}$ and the axial-vector vacuum susceptibility $\chi_{A}$, in order to distinguish them from the later vertex wave function $\chi_{\nu}$, here we use different symbols $\kappa_{V}$ and $\kappa_{A}$ instead. According to the previous derivations, $\kappa_{V}$ can also be written as
\begin{eqnarray}\label{corre-vector1}
\kappa_{V}\langle\bar{q}q\rangle=\int\!\!d^4z\int\!\!\frac{d^4P}{(2\pi)^4}~e^{i P\cdot z}\frac{{\cal{V}}_{\mu }(z){\cal{V}}_{\nu}(z)}{{\cal{V}}^{2}(z)}\left[\Pi^{V}_{\mu\nu}(P) -\Pi^{V,PT}_{\mu\nu}(P)\right] \, ,
\end{eqnarray}
where $\Pi^{V}_{\mu\nu}(P)$ is the vector vacuum polarization
\begin{equation}\label{corre-vector}
\Pi^{V}_{\mu\nu}(P)=-Z_{2}\int^{\Lambda}_q tr\left[\gamma_{\mu}\chi_{\nu}(q;P)\right],
\end{equation}
while for $\kappa_{A}$,
\begin{eqnarray}\label{corre-axial1}
\kappa_{A}\langle\bar{q}q\rangle =\int d^4z\int \frac{d^4P}{(2\pi)^4}~e^{i P\cdot z}\frac{{\cal{V}}_{5\mu}(z){\cal{V}}_{5\nu}(z)}{{\cal{V}}_5^{2}(z)}\left[\Pi^{A}_{\mu\nu}(P)-\Pi^{A,PT}_{\mu\nu}(P)\right]\, ,
\end{eqnarray}
with $\Pi^{A}_{\mu\nu}(P)$ the axial-vector vacuum polarization
\begin{equation}\label{corre-axial}
\Pi^{A}_{\mu\nu}(P)=-Z_{2}\int^{\Lambda}_qtr\left[\gamma_{5}\gamma_{\mu}\chi_{5\nu}(q;P)\right],
\end{equation}
in which the trace is to be taken in color and Dirac space. ${\cal{V}}_{\mu}(z)$ and ${\cal{V}}_{5\mu }(z)$ represent the variable vector and axial-vector external fields, $\chi_{\nu},\chi_{5\nu}$ are the exact vector and axial-vector vertex wave functions, which can be written as $G\Gamma G$ using the no external field versions of the full quark propagator $G$ and the corresponding vertex functions $\Gamma$. $Z_{2}$ is the renormalization constant of the quark wave function. According to the WTI, the vertex renormalization constant is also equal to $Z_{2}$. $\langle\bar{q}q\rangle$ is he chiral quark condensate.

As can be seen from Eqs.~(\ref{corre-vector1}) and (\ref{corre-axial1}), in order to obtain the non-perturbative vector and axial-vector vacuum susceptibilities, we should subtract $\Pi^{V,PT}_{\mu\nu}(P)$ and $\Pi^{A,PT}_{\mu\nu}(P)$ respectively, which arises from the perturbative effects. In other words, when calculating the vacuum susceptibilities, the mean values of the perturbative vacuum should be subtracted. This is very important and is deserving of additional attention. It is well known that the separation of the perturbative and the non-perturbative contributions from the mean values of the vacuum is somewhat arbitrary. Usually, this arbitrariness is avoided by introducing some normalization point~\cite{NPB.249.445--471}. In such a formula, the condensates will depend on the choice of the normalization point. There are also other methods besides this one, for example, in studies of the mixed quark-gluon condensate, the authors of Ref.~\cite{PRD.67.074004} identified the perturbative vacuum with the Wigner vacuum, since both of them are trivial in the sense that there are no chiral symmetry breaking and confinement, in contrast to Nambu-Goldstone vacuum (the non-trivial vacuum) which corresponds to DCSB (more details can be found in Ref.~\cite{PRD.67.074004}). In Refs.~\cite{PRC.72.035202,PRC.73.035206,PLB.639.248--257}, the authors adopted {the viewpoint of Ref.~\cite{PRD.67.074004}} to calculate the vector, the axial-vector, and the tensor vacuum susceptibilities in the framework of rainbow-ladder approximation of the DSEs approach. For example, in the calculation of the vector vacuum susceptibility, $\Pi_{\mu\nu}^V(P)$ in Eq.~(\ref{corre-vector1}) is calculated in the Nambu-Goldstone vacuum configuration, while $\Pi_{\mu\nu}^{V,PT}(P)$ is calculated in the Wigner vacuum configuration. It is obvious that this calculation depends on the rainbow-ladder approximation of the DSEs approach. {In the literature, there are few theoretical studies related to the vector and axial-vector vacuum susceptibilities, among them, L. S. Kisslinger determined them using a three-point formalism within the method of QCD sum rules~\cite{PRC.59.3377}, M. Harada $et~al$ discussed the effective degrees of freedom at chiral restoration and the vector manifestation in hidden local symmetry theory~\cite{NPA.727.437}, and K. Jo $et~al$ calculated vector susceptibility and QCD phase transition in anti-de Sitter (AdS)/QCD models~\cite{JHEP.11.040}}. In the following, we will show how the authors of Ref.~\cite{PLB.669.327--330} obtain model independent results for the vector and axial-vector vacuum susceptibilities using the vector and axial-vector WTIs.

Let us discuss the vector vacuum susceptibility $\kappa_{V}$ first. The conservation of the vector current ensures that the vector vacuum polarization is purely transverse, namely,
\begin{equation}\label{vector-trans}
\Pi^{V}_{\mu\nu}(P)=\left(\delta_{\mu\nu}-\frac{P_{\mu}P_{\nu}}{P^{2}}\right)\Pi^{V}_{T}(P^{2}).
\end{equation}
If we take the constant external field limit ${\cal V}_{\mu}(z)\equiv{\cal V}_{\mu}$, and use the following integration equation
\begin{equation}\label{integ}
\int d^4 l~\delta^4(l)~\frac{p\cdot l k\cdot l}{l^2}f(k,p,l)=\frac{1}{4}\int d^4 l~\delta^4(l)~p\cdot k f(k,p,l),
\end{equation}
we can get
\begin{eqnarray} \label{sus-begin1}
\kappa_{V}\langle\bar{q}q\rangle=\frac{3}{4}\left[\Pi^{V}_{T}(P^{2}=0) -\Pi^{V,PT}_{T}(P^{2}=0)\right].
\end{eqnarray}
It can be seen that the vector vacuum susceptibility is closely related to the vector vacuum polarization at zero total momentum. Now contracting both sides of Eq.~(\ref{corre-vector}) with $\delta_{\mu\nu}$, we then have
\begin{equation}\label{vector-zero}
\Pi_{T}^{V}(P^{2}=0)=-\frac{Z_{2}}{3}\int^{\Lambda}_{q} tr\left[\gamma_{\mu}\chi_{\mu}(q;P=0) \right].
\end{equation}
The case for $\Pi_{T}^{V,PT}(P^{2}=0)$ is similar.

In the chiral limit case, the wave function of the vector vertex satisfy the WTI
\begin{equation}\label{WTI-vector}
iP_{\mu}\chi_{\mu}(q;P)=G(q_{-})-G(q_{+}),
\end{equation}
where $q_{\pm}=q\pm P/2$. Expanding the right side of Eq.~(\ref{WTI-vector}) to $P_{\mu}$ and taking the limit $P_{\mu}\rightarrow 0$ leads to the Ward identity
\begin{equation}\label{Ward-vector}
\chi_{\mu}(q;P=0)=i\frac{\partial G(q)}{\partial q_{\mu}}. \,
\end{equation}
Substituting Eq.~(\ref{Ward-vector}) into the right side of Eq.~(\ref{vector-zero}), and adopting the following parametrization of the quark propagator $G(q)$
\begin{equation}
\label{Spform} G(q) = \frac{1}{i\gamma\cdot k A(q^2) + B(q^2)} = -i\gamma\cdot k \sigma_{V}(q^2)+ \sigma_{S}(q^2) \, ,
\end{equation}
we obtain
\begin{eqnarray}\label{v-zero-r}
 \Pi_{T}^{V}(P^{2}=0) &=&-\frac{Z_{2}}{3}\int_q^{\Lambda}tr\left[i\gamma_{\mu}\frac{\partial G(q)}{\partial q_{\mu}} \right] \nonumber\\
&=&-8Z_{2}\int_q^{\Lambda}[2\sigma_V(q^2)+q^2\frac{d\sigma_V(q^2)}{d q^2}].
\end{eqnarray}
Since for large $q^2$, $\sigma_V\sim q^{-2}$, this integral is quadratically divergent. However, this divergence is not genuine. Note that the integrand is a total divergence, so the above integral vanishes if a translation invariant regularization is adopted,\footnote{Actually, in numerical calculations of the vacuum polarization using DSEs people usually employ a cutoff to regularize the ultraviolet divergence in Eq.~(\ref{corre-vector}), for the details please see Ref.~\cite{PLB.669.327--330} and references therein.} and hence
\begin{equation}\label{vector-relation}
\Pi_{T}^{V}(P^{2}=0)=0.
\end{equation}
For similar reasons we can get $\Pi_{T}^{V,PT}(P^{2}=0)=0$ too. Thus we may draw the conclusion that the vector vacuum susceptibility is zero so long as the Ward identity is satisfied. In other words, the vanishing of the vector vacuum susceptibility is a direct consequence of gauge invariance. The authors of Ref.~\cite{PRC.72.035202} also made use of the vector Ward identity in constructing the vertex function, but at that time they did not realize that the vector vacuum susceptibility can be expressed with such a simple formula, but their numerical result (of the order $10^{-5}$) is very close to this model independent one.

The case for the axial-vector vacuum polarization is a little bit more complicated, {since the axial-vector vertex contains a massless bound state pole}. Nevertheless, the analysis is still straightforward, and the general form of the axial-vector vacuum polarization can be written as
\begin{equation}\label{axial-trans}
\Pi^{A}_{\mu\nu}(P)=\left(\delta_{\mu\nu} -\frac{P_{\mu}P_{\nu}}{P^{2}}\right)\Pi^{A}_{T}(P^{2})-\frac{P_{\mu}P_{\nu}}{P^{2}}\Pi^A_L (P^2),
\end{equation}
which contains both a transverse part and a longitudinal part. Following the same procedure of deducing the vector vacuum susceptibility, we can get
\begin{equation} \label{sus-begin2}
\kappa_{A}\langle\bar{q}q\rangle  =\frac{3}{4}\left[\Pi^{A}_{T}(P^{2}=0)-\Pi^{A,PT}_{T}(P^{2}=0)\right]-\frac{1}{4}\left[\Pi^{A}_{L}(P^{2}=0)-\Pi^{A,PT}_{L}(P^{2}=0)\right].
\end{equation}
From Eqs.~(\ref{axial-trans}) and (\ref{corre-axial}), we know that
\begin{equation}\label{axial-component-equality}
\left(\delta_{\mu\nu} -\frac{P_{\mu}P_{\nu}}{P^{2}}\right)\Pi^{A}_{T}(P^{2})-\frac{P_{\mu}P_{\nu}}{P^{2}}\Pi^A_L (P^2) =-Z_{2}\int^{\Lambda}_qtr\left[\gamma_{5}\gamma_{\mu}\chi_{5\nu}(q;P)\right].
\end{equation}

Now we will determine the axial-vector vacuum polarization at $P^2=0$. In the chiral limit, the WTI for the axial-vector vertex can be expressed as
\begin{equation}\label{WTI-axial}
-iP_{\mu}\chi_{5\mu}(q;P)=G(q_{+})\gamma_{5}+\gamma_{5}G(q_{-}) \, .
\end{equation}
where we also express the identity in terms of the dressed axial-vector vertex wave function $\chi_{5\mu}$ instead of the dressed vertex $\Gamma_{5\mu}$. Since $\chi_{5\mu}$ possesses a longitudinal massless bound state pole, its generally expression is then~\cite{PLB.420.267--273}
\begin{equation}\label{axial-vertex-ge}
\chi_{5\mu}(q;P)=\gamma_{5}\chi_{\mu}^{R}(q;P)+\tilde\chi_{5\mu}(q;P)+\frac{f_{\pi}P_{\mu}}{P^{2}}\chi_{\pi}(q;P) \,.
\end{equation}
where $\chi_{\mu}^{R}(q;P)=\gamma_{\mu}F_{R}+\gamma\cdot q q_{\mu}G_{R}-\sigma_{\mu\nu}q_{\nu}H_{R}$, functions $F_{R},G_{R},H_{R}$ are regular at $P^{2}\rightarrow 0$, $P_{\mu}\tilde\chi_{5\mu}\sim O(P^{2})$, $f_\pi$ is the
pion decay constant in the chiral limit, and $\chi_{\pi}(q;P)$ is the canonically normalized Bethe-Salpeter wave functions of the massless bound state which take a general form
\begin{equation}
\chi_{\pi}(q;P)  = 2\gamma_5\left[ i E_{\pi}(q;P) + \gamma\cdot PF_{\pi}(q;P)  +\gamma\! \cdot \! q q\!\cdot\! P G_{\pi}(q;P)+\sigma_{\mu\nu}q_\mu P_\nu H_{\pi}(q;P)\right].
\end{equation}
In the chiral limit, $f_{\pi}$ is defined as
\begin{equation}\label{f-pion-g}
f_{\pi}P_{\mu}=Z_{2}\int_{q}^{\Lambda} tr\left[\gamma_{5}\gamma_{\mu}\chi_{\pi}(q;P) \right].
\end{equation}
It is apparent that as $P\rightarrow0$, we can expand the right side of Eq.~(\ref{f-pion-g}) to $O(P_{\mu})$, and then
\begin{equation}\label{f-pion-limit}
f_{\pi}=-24Z_{2}\int_{q}^{\Lambda}\left(F_{\pi}(q;0)+\frac{1}{4}q^{2}G_{\pi}(q;0)\right ).
\end{equation}

Contracting both sides of Eq.~(\ref{axial-component-equality}) with the projector ${\cal P}_{\mu\nu}^4 = \delta_{\mu\nu} - 4\frac{P_\mu P_\nu}{P^2}$, we can get
\begin{equation}\label{projection-4}
3\Pi_T^A(P^2)+3\Pi_L^A (P^2)=-Z_2\int_{q}^{\Lambda} tr\bigg[{\cal P}_{\mu\nu}^4\gamma_5\gamma_\mu(\gamma_5\chi_\nu^R(q;P)+{\tilde\chi}_{5\nu}(q;P)+\frac{f_\pi P_\mu}{P^2}\chi_\pi(q;P))\bigg].
\end{equation}
Not let us focus on the limit of Eq.~(\ref{projection-4}) when $P^2\rightarrow 0$. After some algebra we find that, the first term in the right side of Eq.~(\ref{projection-4}) vanishes, and the second term also vanishes thanks to $P_{\mu}\tilde\chi_{5\mu}\sim O(P^{2})$, while the third term is found to be $-24f_\pi N_c Z_2\int_{q}^{\Lambda}(F_\pi(q;0)+\frac{1}{4}q^2G_\pi(q;0))$, which equals $3f_\pi^2$ after making use of Eq.~(\ref{f-pion-limit}). Therefore we obtain
\begin{equation}\label{transverse-longitudinal-relation}
\Pi_T^A (P^2=0)+\Pi_L^A (P^2=0)=f_\pi^2.
\end{equation}
Similarly, contracting both sides of Eq.~(\ref{axial-component-equality}) with the projector ${\cal P}_{\mu\nu}=\delta_{\mu\nu}-\zeta\frac{P_\mu P_\nu}{P^2}$($\zeta \not=4$) gives
\begin{equation}\label{projection-general}
3\Pi_T^A(P^2)+(\zeta-1)\Pi_L^A(P^2)=-Z_2\int_{q}^{\Lambda}tr\bigg[{\cal P}_{\mu\nu}\gamma_5\gamma_\mu(\gamma_5\chi_\nu^R(q;P)+{\tilde\chi}_{5\nu}(q;P)+\frac{f_\pi P_\mu}{P^2}\chi_\pi(q;P))\bigg].
\end{equation}
In the limit $P^2\rightarrow 0$, the first term of Eq.~(\ref{projection-general}) is $(4-\zeta)4N_c Z_2\int_{q}^{\Lambda}(F_R(q;0)+\frac{1}{4}q^2G_R(q;0))$, the second term does not contribute, while the third term is $(\zeta-1 )f_\pi^2$. Now using Eq.~(\ref{transverse-longitudinal-relation}) we then get
\begin{equation}\label{transverse-relation}
\Pi_T^A (P^2=0)=4N_c Z_2 \int_q^\Lambda [F_R(q;0)+\frac{1}{4}q^2G_R(q;0)].
\end{equation}
Note that the result is independent of $\zeta$.

Now substituting the general form of $\chi_{5\mu}$ in Eq.~(\ref{axial-vertex-ge}) into Eq.~(\ref{WTI-axial}), and expand both sides to $O(1)$ and $O(P_{\mu})$ respectively, we can get the corresponding Goldberger-Treiman relation. Functions $F_{R}(q;0),G_{R}(q;0)$ can be expressed using the vector part of the quark propagator and the BS wave function of pion (details can be found in Ref.~\cite{PLB.420.267--273})
\begin{eqnarray}\label{fwti}
 F_R(q;0)&=&-\sigma_{V}(q^2)-2 f_{\pi}F_{\pi}(q;0),\nonumber\\
 G_R(q;0)&=&-2\frac{d\sigma_{V}(q^2)}{d q^2}-2 f_{\pi}G_{\pi}(q;0).
\end{eqnarray}

Substituting Eq.~(\ref{fwti}) into Eq.~(\ref{transverse-relation}), and using Eq.~(\ref{f-pion-limit}), we get
\begin{eqnarray}\label{transverse-relation-final}
\Pi_T^A(P^2=0)=-2N_c Z_2\int_q^\Lambda[2\sigma_V(q^2)+q^2\frac{d\sigma_V(q^2)}{d q^2}]+f_\pi^2=f_\pi^2.
\end{eqnarray}
To obtain this result we have used the previous discussion that when using a translation invariant regularization, the integral to $q$ in Eq.~(\ref{transverse-relation-final}) is $0$.

Combining Eqs.~(\ref{transverse-longitudinal-relation}) and (\ref{transverse-relation-final}), we have
\begin{equation}\label{axial-component-zero-momentum}
\Pi_T^A(P^2=0)=f_\pi^2,~~~\Pi_L^A(P^2=0)=0.
\end{equation}
The subtraction term $\Pi_{T}^{A,PT}(P^{2}=0)$ and $\Pi_{L}^{A,PT}(P^{2}=0)$ can be obtained using similar methods. Note that the perturbative axial-vector vertex function has no pion pole term. We obtain
\begin{equation}\label{axial-component-zero-momentum-PT}
\Pi_T^{A,PT}(P^2=0)=0,~~~\Pi_L^{A,PT}(P^2=0)=0.
\end{equation}
The axial-vector vacuum susceptibility is then
\begin{eqnarray}\label{final}
\kappa_{A}\langle\bar{q}q\rangle =\frac{3}{4}f_{\pi}^{2}.
\end{eqnarray}

From Eqs.~(\ref{vector-relation}) and (\ref{axial-component-zero-momentum}), we found that in the chiral limit case
\begin{equation}\label{V-A}
\Pi_T^V(P^2=0)-\Pi_T^A(P^2=0)=-f_\pi^2.
\end{equation}
which is just the result from Weinberg sum rules~\cite{PRL.18.507}.

In summary, in this part we have derived model independent results of the vector and the axial-vector vacuum susceptibilities, which will play an important role in the related calculations of the QCD sum rules external
field method. Using the vector and the axial-vector WTIs, Ref.~\cite{PLB.669.327--330} has demonstrated that in the chiral limit the vector vacuum susceptibility is $0$, while the axial-vector vacuum susceptibility equals $3/4$ of the square of the pion decay constant, which is the same as the result from Weinberg sum rules.

\subsubsection{The scalar vacuum susceptibility}\label{scalarsus}
Studies on the QCD phase transitions play important roles in the studies of the evolution of the early universe and the high energy heavy ion collisions. The results of the Lattice QCD show that the QCD ``phase transition'' in the early universe is not a real phase transition, but a smooth crossover~\cite{Nature.443.675--678}. In Ref.~\cite{Nature.443.675--678} they adopted the scalar vacuum susceptibility $\chi$ as an indicator, and studied the temperature dependence of $\chi$. The scalar vacuum susceptibility is very important in studies of QCD phase transitions, however, it is known that there is a square divergence in the $\chi(T)$ calculation. In order to eliminate it, a usual treatment is to subtract $\chi(T=0)$ from $\chi(T)$, since the divergence only exists in $\chi(T=0)$, and the temperature dependent part is not divergent~\cite{PLB.643.46--54,Nature.443.675--678}. Obviously, such method is not applicable in the zero temperature case, but the scalar vacuum susceptibility is meaningful at zero temperature. We will give a new treatment to the square divergence in $\chi(T)$ later.

In QCD, the two-quark condensate associated with $f$ flavor of quarks is defined as~\cite{PRC.67.065206}
\begin{equation}\label{sigmaf}
 \langle\bar{q}q\rangle_f(m_f;\mu,\Lambda) =Z_4(\mu,\Lambda) N_c {\rm tr}_{\rm D} \! \int_q^\Lambda\!G_f(q;m_f;\mu)\,,
\end{equation}
where $m_f(\mu)$ is the renormalized current quark mass, with $\mu$ the renormalization scale; $Z_4$ is the renormalization constant of the mass term in the Lagrangian, which depends implicitly on the gauge parameter. In the chiral limit case the two-quark condensate can be used as an order parameter to describe the chiral phase transition, the scalar vacuum susceptibility (or chiral susceptibility) just characterize its response to the change of quark mass~\cite{PRD.54.1087,PRD.67.114004}
\begin{eqnarray}\label{chif}
\chi(\mu):=\left.\frac{\partial}{\partial m(\mu)}\,\langle\bar{q}q\rangle(m;\mu,\Lambda)\right|_{\hat m=0}=Z_4(\mu,\Lambda)N_c{\rm tr}_{\rm D}\!\int_q^\Lambda\!\left.\frac{\partial}{\partial m}G(q;m;\mu)\right|_{\hat m=0}\,,
\end{eqnarray}
here ${\hat m}$ is the auxiliary current quark mass that can be regarded as renormalization invariant. The order of integration and differentiation can be interchanged because the theory is properly regularized. Now use the Ward identity
\begin{equation}\label{scalarWI}
\frac{\partial}{\partial m} G(q;m;\mu)=-G(q;m;\mu)\Gamma(q,0;\mu)G(q;m;\mu)
\end{equation}
then
\begin{equation}
\chi(\mu)=-Z_4(\mu,\Lambda)N_c{\rm tr}_{\rm D}\!\int_q^\Lambda\!G(q;0;\mu)\Gamma(q,0;\mu)G(q;0;\mu)\,, \label{chim}
\end{equation}
where $\Gamma$ is the renormalized fully-dressed scalar vertex, which satisfies the inhomogeneous BSE
\begin{equation}
\Gamma(k,P;\mu) = Z_4 \mbox{\boldmath $I_D$}+ \int_q^\Lambda [S(q_+)\Gamma_0 (q,P) S(q_-)]_{sr} K_{tu}^{rs}(q,k;P). \label{scalarBSE}
\end{equation}
with $k$ the relative and $P$ the total momenta of the quark--anti-quark pair; $q_\pm = q\pm P/2$; $r,s,t,u$ represent color and Dirac indices; and $K$ is the fully-amputated quark--anti-quark scattering matrix.

The quantity $m^2\Pi_0(P)$ is renormalization point invariant in QCD, where $\Pi_0(P)$ is the scalar vacuum polarization
\begin{equation}
\Pi_0(P;\mu) = Z_4 \, N_c{\rm tr}_{D} \int_q^\Lambda\! G(q_+)\Gamma(q,P) G(q_-)\,.
\end{equation}
To compare with Eq.~(\ref{chim}), we can get the general result
\begin{equation}
\label{chimpol} \chi(\mu) = - \Pi_0(P=0;\hat m=0,\mu) \,.
\end{equation}

Hitherto we have not specified a regularization procedure for the scalar vacuum susceptibility. In this connection it is noteworthy that if a hard cutoff is used, then in the chiral limit of a non-interacting theory
\begin{equation}\label{Lambdadiverge}
-\Pi_0^0(P=0;\hat m=0)=\frac{N_c}{4\pi^2}\Lambda^2.
\end{equation}
Ref.~\cite{PRC.67.065206} pointed out that, this result can be traced to the dependence on current quark mass in Eq.~(\ref{sigmaf}). On the other hand, the Pauli-Villars regularization would yield zero, which is then the procedure that the authors of Ref.~\cite{PRC.79.035209} recommend and employ in models that preserve the one-loop renormalization-group behavior of QCD, and in this case the scalar vacuum susceptibility is defined self-consistent.

For the calculation of $\chi(\mu)$ we only need the scalar vertex at $P=0$, at which total momentum it has the general form
\begin{equation}\label{candd}
  \Gamma_0(k,0;m)=i\gamma\cdot k C(k^2;m)+D(k^2;m).
\end{equation}
Owing to the Ward identity, Eq.~(\ref{scalarWI}), it is not necessary to solve the inhomogeneous BSE since
\begin{equation}\label{cadb}
C(k^2;m)=\frac{\partial}{\partial m}A(k^2;m);~~~~~D(k^2;m)=\frac{\partial}{\partial m}B(k^2;m).
\end{equation}
Hence a solution of the gap equation suffices completely to fix $\Gamma_0(k,0)$.

A simplified form of the renormalization group improved effective interaction is employed in Ref.~\cite{PRC.79.035209}; viz., they retain only that piece which expresses the long-range behavior~\cite{PRC.56.3369,PRC.60.055214}
\begin{equation}\label{IRGs}
g^2D_{\mu\nu}(q)=4\pi^2t_{\mu\nu}(q)D\frac{q^2}{\omega^6}\exp(-\frac{q^2}{\omega^2}),
\end{equation}
with $t_{\mu\nu}(q)$ the transverse momentum projection operator. Eq.~(\ref{IRGs}) delivers an ultraviolet finite model gap equation. Hence, the regularization mass scale $\Lambda$ can be moved to infinity and the renormalization constants equal to one.

The active parameters in Eq.~(\ref{IRGs}) are $D$ and $\omega$, but they are not independent. In reconsidering a renormalization group improved rainbow-ladder fit to a selection of ground state observables~\cite{PRC.60.055214}, Ref.~\cite{EPJA.18.231-235} noted that a change in $D$ is compensated by an alteration of $\omega$. This feature has further been elucidated and exploited in Refs.~\cite{PRC.77.042202,PRC.79.012202}. For the range $[0.3,0.5]\,$GeV, the fitted low-energy observables are approximately constant along the trajectory
\begin{equation}\label{gluonmass}
\omega D  = (0.8 \, {\rm GeV})^3 =: m_g^3\,
\end{equation}
Herein, we employ $\omega=0.5\,$GeV, and therefore $D=m_g^3/\omega=1.0\,$GeV$^2$ corresponds to what might be called the real-world reference value for the bare vertex. Here we should note that, if we choose Eq.~(\ref{IRGs}) as the model gluon propagator, but for the quark-gluon vertex using the BC ansatz Eq.~(\ref{bcvertex}), there is no reason to suppose that $\omega$ and $D$ still satisfy the relation in Eq.~(\ref{gluonmass}), since the BC vertex contains considerable unknown non-perturbative information of interaction. Therefore, we need to refit the parameter $D$. After some tries, it has been found that if one assumes $D=0.5\, $GeV$^2$, we can get the value of the two-quark condensate shown in the fourth column of Table~\ref{Table:Para1}. Hence we postulate that $D=0.5\, $GeV$^2$ is the real world reference value under the BC vertex.
\begin{table}[htb]
\caption{Results obtained for selected quantities with $\omega=0.5\,$GeV, and the vertex and $D$ parameter value indicated: $A(0)$, $M(0)$ are $p=0$ in-vacuum values of the scalar functions $A(p^2)$ and $M(p^2)=B(p^2)/A(p^2)$; the quark condensate is defined with $m=0$ in Eq.~(\ref{sigmaf}); and $\chi$ is obtained from Eq.~(\protect\ref{chipractical}). A(0) is dimensionless, but all other entries are quoted in GeV. The calculations reported herein were performed in the chiral limit, and all are taken from Ref.~\cite{PRC.79.035209}}
\centering
\begin{tabular}{@{}cccccc@{}}\hline
  Vertex & $\sqrt D$ & $A(0)$ & $M(0)$ & $-(\langle\bar q q \rangle^0)^{1/3}$ & $\sqrt\chi$\\\hline
  {\rm Rainbow-Ladder} &1 & 1.3 & 0.40 & 0.25 & 0.39\\\hline
  {\rm Ball-Chiu} & $1/\sqrt2$ & 1.1 &0.28 & 0.26 & 0.28 \\\hline
\end{tabular}\label{Table:Para1}
\end{table}

As discussed above, Eq.~(\ref{Lambdadiverge}) demonstrates that a regularization procedure is needed for the vacuum polarization, even if dealing with a free field theory. For QCD, in which Schwinger functions carry anomalous dimensions, the Pauli-Villars scheme works well. However, this procedure fails when the vacuum polarization is evaluated with Schwinger functions generated by Eq.~(\ref{IRGs}), where the interaction does not preserve the one-loop behavior of QCD, but delivers an ultraviolet finite gap equation instead. In the following we will construct a simple alternative.

The vector vacuum polarization is given by
\begin{equation}
\Pi^V_{\mu \nu}(P) = N_c {\rm tr}_{\rm D} \! \int_q^\Lambda\! i\gamma_\mu G(q_+) i\Gamma_\nu (q,P) G(q_-)\,,
\end{equation}
here $\Gamma_\nu$ is the vector quark--anti-quark vertex that satisfies an inhomogeneous BSE similar to Eq.~(\ref{scalarBSE}). This polarization couples to the photon and hence must be transverse while contain no mass term. Consider therefore~\cite{PLB.669.327--330}
\begin{eqnarray}
\frac{1}{4}\Pi^V_{\mu \mu}(P=0) &=& \frac{N_c}{4} {\rm tr}_{\rm D} \int_q^\Lambda i \gamma_\mu G(q) i\Gamma_\mu (q,0) G(q) \nonumber \\
&=&-\frac{N_c}{4} {\rm tr}_{\rm D}  \int_q^\Lambda i \gamma_\mu\frac{\partial}{\partial q_\mu} G(q) \rule{3em}{0ex} \nonumber \\
&=&-2 N_c \int_q^\Lambda \frac{1}{q^2} \frac{d}{dq^2}\left[ (q^2)^2\sigma_V(q^2) \right],\rule{3em}{0ex} \label{vectorvacuum}
\end{eqnarray}
where the vector Ward identity was used in the second line and we have written the dressed quark propagator in the following form
\begin{equation}
G(p) = -i \gamma\cdot p \, \sigma_V(p^2) + \sigma_S(p^2)\,.
\end{equation}
If we consider a free field and use a hard cutoff, then Eq.~(\ref{vectorvacuum}) becomes
\begin{equation}
- \frac{1}{4}\Pi^{V 0}_{\mu \mu}(P=0) = \frac{N_c}{8 \pi^2}\Lambda^2;
\end{equation}
namely, the same sort of divergence is encountered as in the scalar vacuum polarization, Eq.~(\ref{Lambdadiverge}). Naturally, any regularization scheme that preserves the vector Ward-Takahashi would yield zero as the result~\cite{PRD.46.2695}. So, considering Eq.~(\ref{vectorvacuum}), we can define the regularized scalar vacuum susceptibility as
\begin{eqnarray}\label{chipractical}
\lefteqn{\chi=-\Pi_0(0;m=0,\Lambda)+\frac{1}{2}\Pi^{V}_{\mu\mu}(0;m=0,\Lambda)} \\\nonumber
 &=&-\frac{N_c}{4\pi^2}\int_0^\infty ds\,s\,\bigg\{D(s)[\sigma_s(s)^2-s\sigma_s(s)^2] \\\nonumber
 &&+2s\, C(s)\sigma_V(s)\sigma_S(s)+2\sigma_V(s)+s\sigma_V^\prime(s)\bigg\}.
\end{eqnarray}
Here, as long as each term in Eq.~(\ref{chipractical}) is regularized independently in a valid fashion, then this is a ``nugatory'' transformation: since the photon is massless and hence in any valid scheme there must be $\Pi^{V}_{\mu \mu}(0;0,\Lambda)=0$. In other words, with this definition we can move on to calculate the susceptibility without specifying a particular regularization scheme. The last two contributions in the integrand act as the Pauli-Villars terms, and $Z_4=1$ in the ultraviolet-finite model.

\begin{figure}[htb]
\centering
\includegraphics[width=0.49\textwidth]{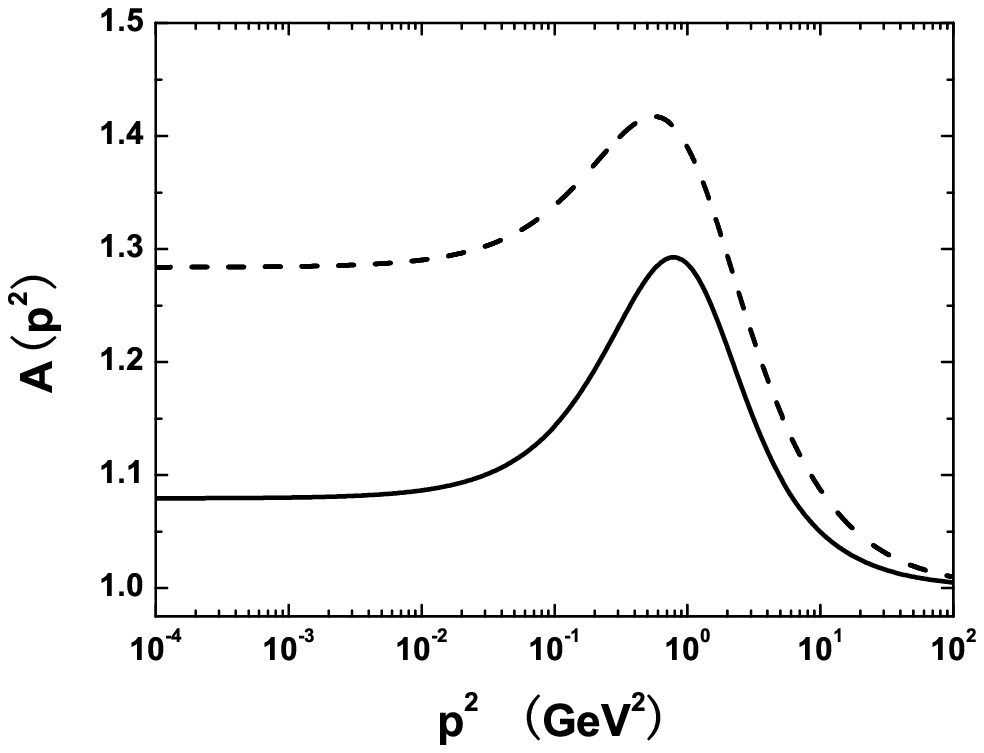}
\includegraphics[width=0.49\textwidth]{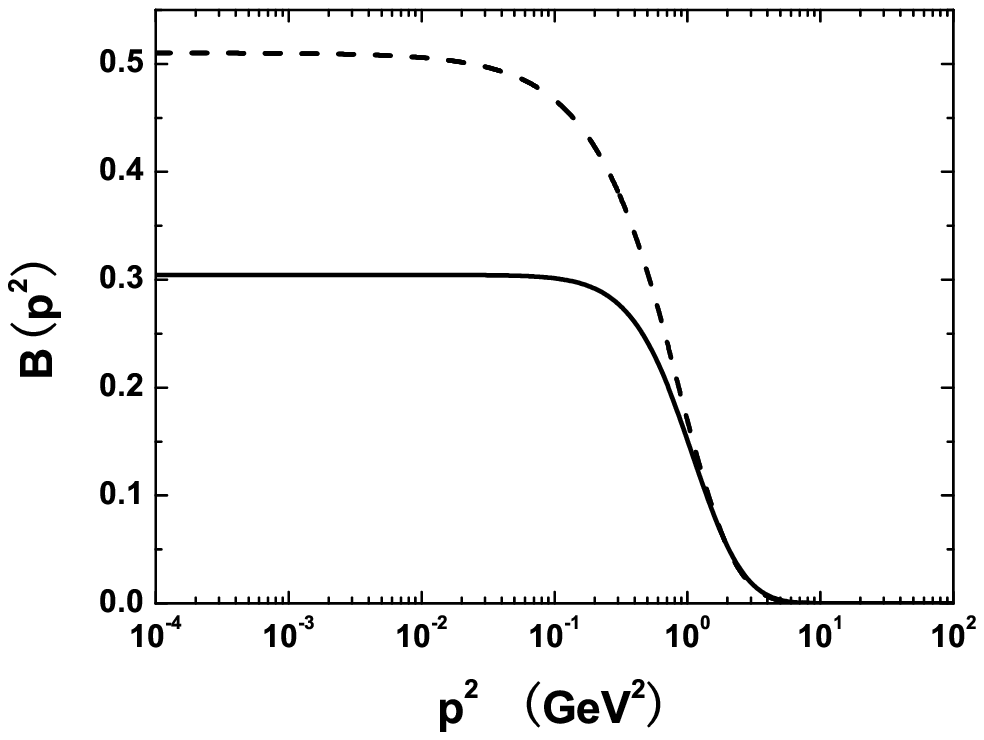}
\caption{\label{aamm} The dressed quark propagator, taken from Ref.~\cite{PRC.79.035209}. Left panel: $A(p^2)=1/Z(p^2)$, where $Z(p^2)$ is the wave function renormalization function; Right panel: $B(p^2)$ {[GeV]}, the scalar piece of the dressed-quark self-energy. In both panels, the dashed curve is calculated in the rainbow-ladder truncation, with ${\cal I}=D/\omega^2=4$; while the solid curve is calculated with the BC vertex ansatz, and ${\cal I}=2$.}
\end{figure}

\begin{figure}[htb]
\includegraphics[clip,width=0.49\textwidth]{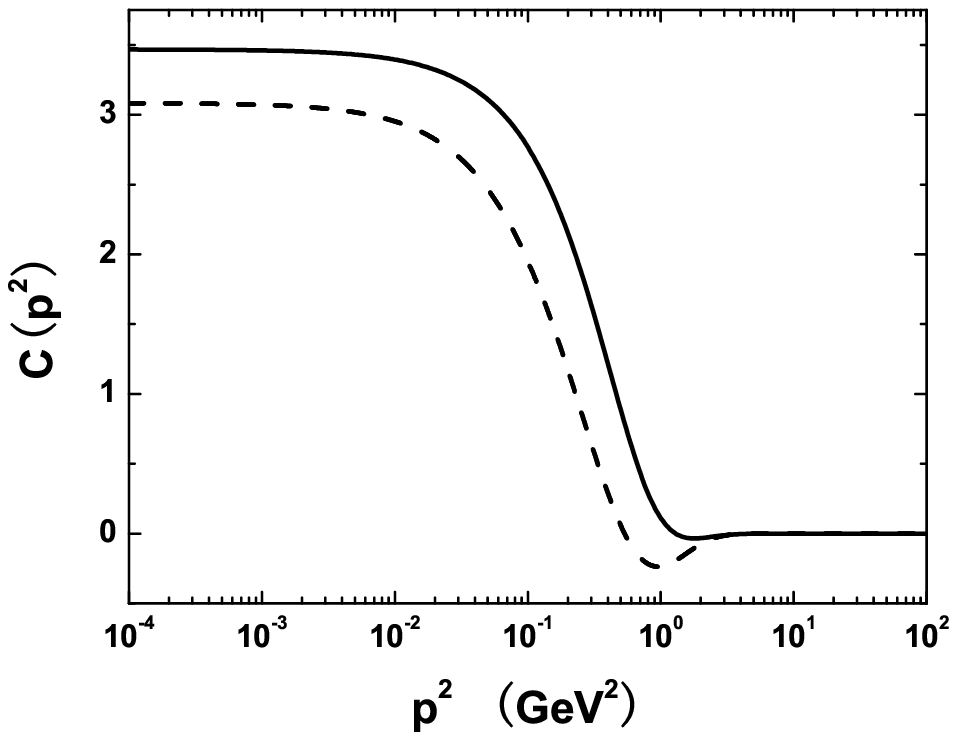}
\includegraphics[clip,width=0.49\textwidth]{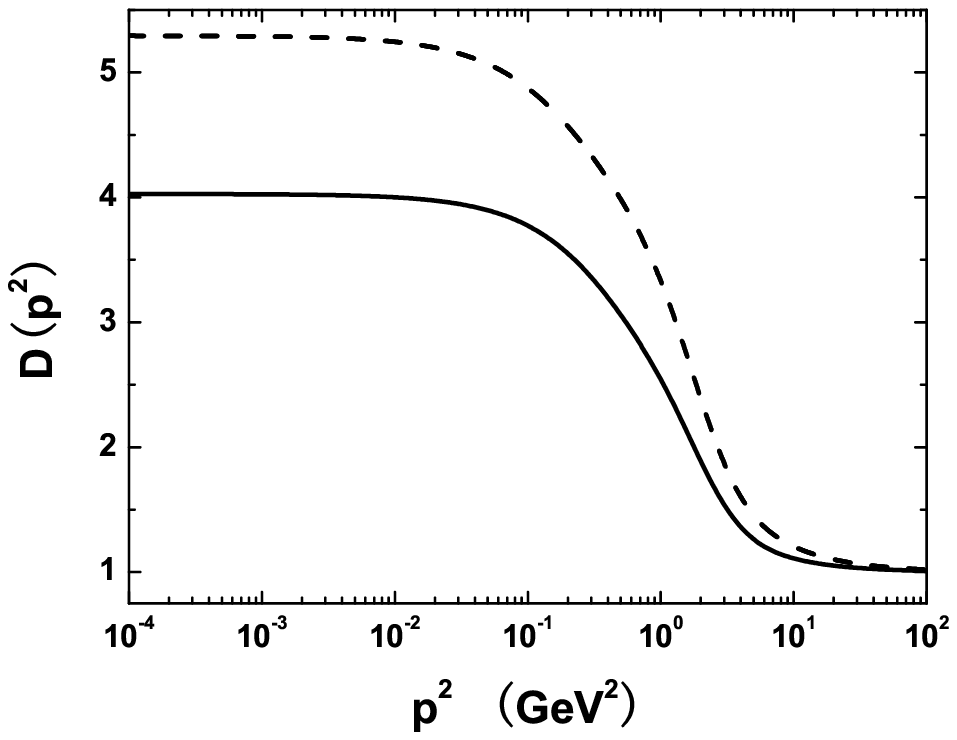}
\caption{\label{ccdd}$P=0$ scalar vertex, taken from Ref.~\cite{PRC.79.035209}. Left panel: $C(p^2)$; Right panel: $D(p^2)$. In both panels, the dashed curve is calculated in the rainbow-ladder truncation, with ${\cal I}=D/\omega^2=4$; while the solid curve is calculated with the BC vertex ansatz, and ${\cal I}=2$.}
\end{figure}

Now we will discuss the numerical results. Owing to the Gaussian form of the gluon propagator, Eq.~(\ref{IRGs}), all the relevant integrations converge rapidly. Fig.~\ref{aamm} shows the functions obtained by solving the gap equation, and Fig.~\ref{ccdd} gives the results which describe the $P=0$ scalar vertex. All the results were obtained with the appropriate real world values of the interaction strength: $D=1\,$GeV$^2$ for the bare vertex, and $D=0.5\,$GeV$^2$ for the BC vertex, as listed in Table~\ref{Table:Para1}.

It can be seen clearly from Fig.~\ref{aamm} that, the BC vertex ansatz has a quantitative impact on the magnitude and pointwise evolution of the gap equation’s solution. This can be anticipated from Ref.~\cite{PLB.285.347--353}. Moreover, the pattern of behavior can be understood from Ref.~\cite{PRC.70.035205}: the feedback arising through the $\Delta_B$ term in the BC vertex always acts to alter the domain upon which $A(p^2)$ and $B(p^2)$ differ significantly in magnitude from their respective free-particle values, especially in the intermediate momentum region. The behaviors of $C(p^2)$ and $D(p^2)$ shown in Fig.~\ref{ccdd} can be understood easily from those of $A(p^2)$ and $B(p^2)$, so that there is no need to discuss it further.

\begin{figure}[htb]
\centering
\includegraphics[clip,width=0.7\textwidth]{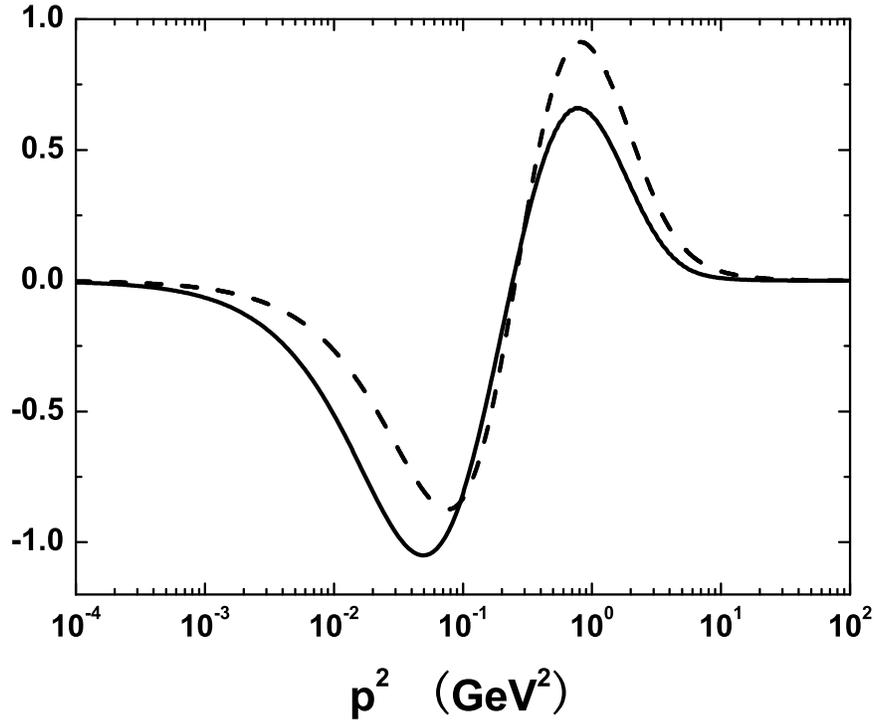}
\caption{\label{integrandRW} Integrand in Eq.~(\protect\ref{chipractical}), taken from Ref.~\cite{PRC.79.035209}. Dashed curve: calculated in rainbow-ladder truncation, with ${\cal I}=D/\omega^2=4$; solid curve: calculated with BC vertex ansatz, and ${\cal I}=2$.}
\end{figure}

The integrand in Eq.~(\ref{chipractical}) is depicted in Fig.~\ref{integrandRW} for each vertex ansatz at the associated real world interaction strength. The resulting scalar vacuum susceptibilities are presented in the fifth column of Table~\ref{Table:Para1}. It can be found from this figure that the regularization has served to eliminate the far-ultraviolet tail of the integrand, thereby ensuring convergence of the integral. We have varied the detailed form of the regularizing subtraction, namely, using free-field propagators and vertices instead of gap and BSE solutions, and the behavior of the integrand changes a little. Moreover, at real world values of the interaction strength for both ansatzs the integrands have negative support in the infrared and positive support for $p^2>(0.5GeV)^2$. These results tell us that when studying the scalar vacuum susceptibility, to go beyond the rainbow-ladder truncation is important and necessary, which then deserves further investigation.

\begin{figure}[htb]
\centering
\includegraphics[clip,width=0.7\textwidth]{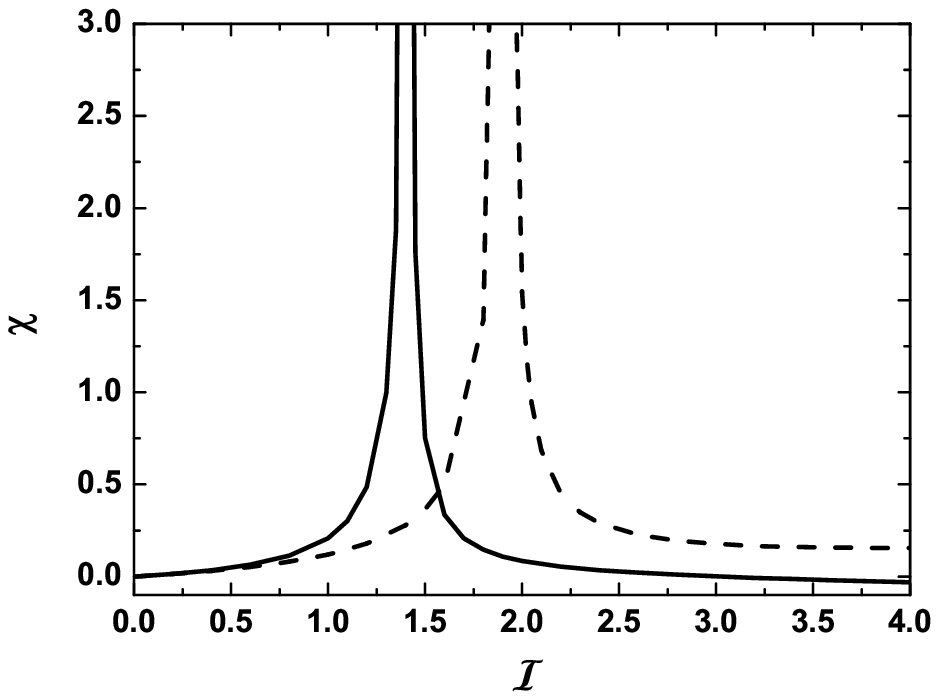}
 \caption{\label{chiI} Dependence of the scalar vacuum susceptibility on the interaction strength in Eq.~(\protect\ref{IRGs}), taken from Ref.~\cite{PRC.79.035209}; viz., ${\cal I}:=D/\omega^2$: dashed curve is the  RL vertex, while solid curve is BC vertex。}
\end{figure}

In Fig.~\ref{chiI} we depict the evolution of the scalar vacuum susceptibility with increasing interaction strength, ${\cal I}=D/\omega^2$. For ${\cal I}=0$ there is no interaction and then the ``vacuum'' is unperturbed by a small change in the current quark mass, hence the susceptibility keeps zero. With increasing ${\cal I}$ the susceptibility will grow since the interaction in the $\bar q q$ channel is attractive and therefore magnifies the associated pairing. This is equivalent to stating that the scalar vertex is enhanced above its free field value. Then the growth continues and accelerates until at some critical value ${\cal I}_c$ the susceptibility becomes infinite, namely, a divergence appear. The critical values are ${\cal I}_c=1.93$ for the rainbow-ladder approximation, and ${\cal I}_c=1.41$ for the BC vertex ansatz.

This divergence can be understood in this way: the models we have defined contain a dimensionless parameter ${\cal I}$, which characterizes the interaction strength, and a current quark mass, which is an explicit source of chiral symmetry breaking. In the general theory of phase transitions, the latter is analogous to an external magnetic field while $1/{\cal I}$ is kindred to a temperature. Consider the free energy for such theories $f(t, m)$, where
$t=[1-{\cal I}_c/{\cal I}]$. If such a theory possesses a second-order phase transition, then the free energy is a homogeneous function of its arguments in the neighborhood of $t=0=m$. From this it follows that the theory's magnetization exhibits the following behavior (e.g., see the Appendix of Ref.~\cite{PRC.58.1758} for the details):
\begin{equation}
M(t,0)=t^\beta,\; t\to 0^+,
\end{equation}
and the associated magnetic susceptibility evolves in this way
\begin{equation}
M(0,m)\propto m^{-(1-1/\delta)}.
\end{equation}
For a mean-field theory, $\beta = 1/2$, and $\delta = 3$.

Now the nature of the critical interaction strength is easy to understand. In the class of theories we are considering, the quark condensate is analogous to the magnetization, and it is attended by the scalar vacuum susceptibility. For ${\cal I}<{\cal I}_c$, the interaction is insufficient to generate a nonzero scalar term in the dressed-quark self-energy in the absence of a current quark mass; namely, DCSB is impossible and the model realizes chiral symmetry in the Wigner mode. But the situation changes at ${\cal I}_c$, and for ${\cal I}>{\cal I}_c$ a $B\neq 0$ solution is always possible. Moreover, the behaviors of the susceptibility show that each model undergoes a second-order phase transition and realizes chiral symmetry in the Nambu-Goldstone mode for interaction strengths above their respective values of ${\cal I}_c$. These observations emphasize the usefulness of the scalar vacuum susceptibility. With the bare vertex people can construct the pressure explicitly,\footnote{The pressure is defined as the negative of the effective action, and then the ground state of a system is that configuration for which the pressure is a global maximum. For the rainbow-ladder approximation, the corresponding pressure is just constructed from the famous CJT effective action~\cite{PRD.10.2428}.} and thereby show that for ${\cal I}>{\cal I}_c$ the DCSB solution is dynamically favored because it corresponds to the configuration of the maximum pressure. Nevertheless, the diagrammatic content of the BC vertex ansatz is not knowable and hence an expression for the pressure cannot be derived. In this case we may rely on the behavior of some susceptibilities to conclude that DCSB is favored, as illustrated within many model studies. Once past the critical point, the susceptibility decreases as ${\cal I}$ increases. This is because the magnitude of the condensate order parameter grows in tandem with ${\cal I}$ and, therefore, the influence of any perturbation associated with a current quark mass must steadily diminish. We will do some related discussions in Sec.~\ref{apps} of this paper. {For the theoretical studies of the scalar and the following pseudo-scalar vacuum susceptibilities using the QCD sum rules, please refer to Ref.~\cite{PRC.59.3377}}.

\subsubsection{The pseudo-scalar vacuum susceptibility}
The color-singlet current-current correlators (or equivalently, the associated vacuum polarizations) are directly related to physical observables, and hence play important roles in QCD. For example, the vector vacuum polarization is coupled to real and virtual photons, so that it is basic to the analysis and understanding of the process $e^+e^-\rightarrow$hadrons~\cite{IJMPA.7.5607--5624,IJMPE.17.870--890}. {The standard Lattice QCD methodology related to the correlators can be found in Ref.~\cite{FBS.40.209--235}}; on the other hand, correlators are also amenable to analysis via the operator product expansion and are therefore fundamental in the application of QCD sum rules. In the latter connection, the pseudo-scalar vacuum susceptibility (or the pion susceptibility) plays a role in the QCD sum rules estimate of numerous meson-hadron couplings, for example, the strong and parity-violating pion-nucleon couplings, $g_{\pi NN}$ and $f_{\pi NN}$, respectively~\cite{PLB.367.21--27,NPB.213.109--121,PRD.57.2847}. Furthermore, the pseudo-scalar vacuum susceptibility is also a probe of QCD vacuum structure as the scalar susceptibility which is discussed in Sec.~\ref{scalarsus} of this paper~\cite{PRC.79.035209}, while its analysis is more subtle, with conflicts and misconceptions being common in different model calculations~\cite{PLB.367.21--27,PLB.576.289--296,EPJA.16.291--297,PRD.57.2847,PLB.136.273--278}.

In this part we derive a model independent result for the pseudo-scalar vacuum susceptibility using the isovector--pseudo-scalar vacuum polarization~\cite{PRC.81.032201}, which can be written as
\begin{equation}\label{ivps}
\Pi_5^{jk}(P;\zeta)=Z_4\,N_c\,{\rm tr}\int_q^\Lambda\frac{i}{2}\gamma_5\tau^j G(q_+)i\Gamma_5^k(q;P) G(q_-)\,.
\end{equation}
where $\zeta$ represents the renormalization scale, the trace is over flavor and spinor indices, {$\tau^j$ is the Pauli matrices of the isospin}, $P$ is the total momentum of the quark--anti-quark pair; $G(q_\pm)$ represents the dressed-quark propagator, and $\Gamma_5$ is the fully dressed pseudo-scalar vertex, both of which depend on the renormalization point. Physical quantities obtained from Eq.~(\ref{ivps}) are then gauge invariant. The propagator can be obtained from the gap equation, Eq.~(\ref{quark-pro}).

The pseudo-scalar vertex is solved from an inhomogeneous BSE, namely
\begin{equation}
[\Gamma_5^j(k;P)]_{tu}=Z_4[\frac{1}{2}\gamma_5\tau^j]_{tu}+\int_q^\Lambda[\chi_5^j(q;P)]_{sr}K_{tu}^{rs}(q,k;P),
\end{equation}
here $k$ is the relative momentum, $r, s, t,$ and $u$ represent color, flavor, and spinor indices, and
\begin{equation}
\chi_5^\mu(k;P)=G(k_+)\Gamma_5^\mu(k;P)G(k_-),
\end{equation}
with $k_\pm=k\pm P/2$, and $K(q,k;P)$ is the fully amputated two-particle irreducible quark--anti-quark scattering kernel. Now, let us consider the case with a space-time independent pseudo-scalar source, $\vec{s}_5\neq0$, associated with the term
\begin{equation}
\int {\rm d}^4x\bar{q}(x)\frac{i}{2}\gamma_5\vec{\tau}\vec{s}_5 q(x)
\end{equation}
in the action, we can define a pseudo-scalar vacuum condensate, the gauge-invariant as well as properly renormalized form of which in QCD is
\begin{equation}
\langle\bar{q}q\rangle_5^\mu(\vec{s}_5,m;\zeta,\Lambda)=Z_4\, N_c{\rm tr}\int_q^\Lambda\frac{i}{2}\gamma_5\tau^\mu G(q;\vec{s}_5,m;\zeta)\,.
\end{equation}
We can see that it is analogous to the vacuum quark condensate, Eq.~(\ref{sigmaf}). When $\hat{m}=0$, the DCSB can be expressed via the Higgs mechanism as
\begin{equation}\label{DCSB0}
-\langle\bar{q}q\rangle_\zeta^0=\lim_{m\rightarrow0}\langle\bar{q}q\rangle(m;\zeta,\Lambda)\neq0.
\end{equation}
Eq.~(\ref{DCSB0}) then defines what we mean by an isoscalar-scalar configuration: isovector--pseudo-scalar correlations are by convention measured with respect to this configuration. These observations highlight the importance of the pseudo-scalar vacuum susceptibility
\begin{equation}\label{pssus}
\chi_5^{\mu\nu}(\zeta)=\left.\frac{\partial}{\partial s_5^\mu}\langle\bar{q}q\rangle_5^\nu(\vec{s}_5,m;\zeta,\Lambda)\right|_{\vec{s}_5=0}.
\end{equation}
Then following the previous discussions, we can know that
\begin{equation}\label{e14}
\chi_5^{\mu\nu}(\zeta)=-2\Pi_5^{\mu\nu}(P=0;\vec{s}_5=0,\hat{m},\zeta).
\end{equation}
Hitherto we have not specified a regularization procedure for the susceptibility, actually it can rigorously be defined via a Pauli-Villars procedure, as discussed in Ref.~\cite{PRC.79.035209}.

Now let us discuss the value of $\chi_5^{\mu\nu}(\zeta)$ in the neighborhood of the chiral limit, therein we may write~\cite{PLB.420.267--273}
\begin{equation}\label{e15}
i\Gamma_5^\mu(k;0)=\frac{1}{2}i\gamma_5\tau^\mu E_5^R(k;0)+\frac{r_\pi}{m_\pi^2}\Gamma_\pi\mu(k;0).
\end{equation}
where $E_5^R(k;P)$ is a part of the inhomogeneous pseudo-scalar vertex that is regular as $P^2+m_\pi^2\rightarrow0$, $\Gamma_\pi\mu(k;P)$ is the pion bound state's canonically normalized Bethe-Salpeter amplitude, and $r_\pi(\zeta)$ determined by
\begin{equation}
 i\delta^{\mu\nu}r_\pi(\zeta)=\langle0|\bar{q}\frac{1}{2}\gamma_5\tau^\nu q|\pi^\mu\rangle =Z_4\,N_c{\rm tr}\int_q^\Lambda\frac{1}{2}\gamma_5\tau^\nu\chi_\pi^\mu(q;P)\,
\end{equation}
is the residue of this bound state in the inhomogeneous pseudo-scalar vertex. Using this, the pion's leptonic decay constant can be expressed as
\begin{equation}
 \delta^{\mu\nu}f_\pi P_\sigma=\langle0|\bar{q}\frac{1}{2}\gamma_5\gamma_\sigma\tau^\nu q|\pi^\mu\rangle =Z_2\,N_c{\rm tr}\int_q^\Lambda\frac{1}{2}\gamma_5\gamma_\sigma\tau^\nu\chi_\pi^\mu(q;P)\,.
\end{equation}

$E_5^R(k;P)$ can be determine via the axial-vector WTI
\begin{equation}
 P_\nu\Gamma_{5\nu}^\mu(k;P)+2m(\zeta)i\Gamma_5^\mu(k;P)=G(k_+)^{-1}\frac{1}{2}i\gamma_5\tau^\mu+\frac{1}{2}i\gamma_5\tau^\mu G(k_-)^{-1}\,,
\end{equation}
with $\Gamma_{5\nu}^\mu$ the inhomogeneous axial-vector vertex. At $P=0$ with $\hat{m}\neq0$ there is no pole contribution on the left side and hence
\begin{equation}
 m(\zeta)E_5^R(k;P=0)=B(k^2;m;\zeta^2)\,.
\end{equation}
in other words, this regular piece of the pseudo-scalar vertex is completely determined by the scalar part of the $\hat{m}\neq0$ quark self-energy. Using the systematic, non-perturbative, and symmetry-preserving DSE truncation scheme introduced in Refs.~\cite{PLB.380.7--12,PRD.52.4736}, we can verify this equation order-by-order via the gap and Bethe-Salpeter equations.

Now substituting Eq.~(\ref{e15}) into Eq.~(\ref{e14}), we then get
\begin{eqnarray}
  \chi_5^{\mu\nu} &\overset{\hat{m}\sim0}{=}& \delta^{\mu\nu}\chi_5(\zeta) \\
  \chi_5(\zeta) &=& \chi_5^\pi(\zeta)+\chi_5^R(\zeta)+O(\hat{m})\,,
\end{eqnarray}
so that, in the neighbourhood of $\hat{m}=0$, the pseudo-scalar susceptibility splits into a sum of two terms: the first one expresses the contribution of the pion pole, $O(\hat{m}^{-1})$, and can be expressed in the closed form
\begin{equation}
\chi_5^\pi(\zeta)=\frac{2r_\pi(\zeta)^2}{m_\pi^2}\overset{\hat{m}=0}{=}-\frac{\langle\bar{q}q\rangle^0_\zeta}{m(\zeta)}\,,
\end{equation}
of which the last equality is proved in Ref.~\cite{PLB.420.267--273};  the second term $O(\hat{m}^0)$ can be determined via
\begin{equation}
m(\zeta)\chi_5^R(\zeta)\delta^{\mu\nu}\overset{\hat{m}\sim0}{=}Z_4\, N_c{\rm tr}\int_q^\Lambda i\gamma_5\tau^\nu G(q)\frac{i}{2}\gamma_5\tau^\mu B(q^2,m)G(q)=\delta^{\mu\nu}\langle\bar{q}q\rangle(m;\zeta,\Lambda)\,,
\end{equation}
where $\langle\bar{q}q\rangle(m;\zeta,\Lambda)$ is the vacuum quark condensate defined in Eq.~(\ref{sigmaf}), and this entails
\begin{equation}
\chi_5^R(\zeta;m)=\chi(\zeta)+O(\hat{m})\,,
\end{equation}
where $\chi(\zeta)$ is the scalar vacuum susceptibility defined in Eq.~(\ref{chif}). Hence we arrive at a model independent consequence of chiral symmetry and the pattern by which it is broken in QCD, namely,
\begin{equation}\label{e28}
 \chi_5(\zeta)\overset{\hat{m}\sim0}{=}-\frac{\langle\bar{q}q\rangle^0_\zeta}{m(\zeta)}+\chi(\zeta)+O(\hat{m})\,.
\end{equation}

Now we show the numerical results of Ref.~\cite{PRC.81.032201} in Table~\ref{psstable}, which were computed from two models for the gap equation’s kernel.
\begin{table}[htb]
\caption{Pseudo-scalar vacuum susceptibility and related quantities computed using the two kernels of the BSE described in connection with Eqs.~(\ref{bcvertex}) and (\ref{IRGs}), also similar to those of Table~\ref{Table:Para1}. Dimensioned quantities are listed in GeV, and all are taken from Ref.~\cite{PRC.79.035209}}
\centering
\begin{tabular}{@{}cccccccc@{}}\hline
  Vertex & $\sqrt D$ & $\omega$ & $-(\langle\bar q q\rangle_\zeta^0)^{1/3}$ & $f_\pi^0$ & $m$ & $\sqrt{\chi_5^\pi}$ & $\sqrt{\chi_5^R}$\\\hline
  {\rm Rainbow-Ladder} &1 & 0.5 & 0.25 & 0.091 & 0.0050 & 1.77& 0.39\\\hline
  {\rm Ball-Chiu} & $1/\sqrt2$ & 0.5 &0.26 & 0.11 & 0.0064&1.66&0.28 \\\hline
\end{tabular}\label{psstable}
\end{table}

Eq.~(\ref{e28}) is a meaningful result. Recall that, in the absence of a current-quark mass, the two-flavor action has a SU$_L(2)\bigotimes$ SU$_R$(2) symmetry and, moreover, that ascribing scalar-isoscalar quantum numbers to the QCD vacuum is a convention contingent upon the form of the current-quark mass term. It follows that, the massless action cannot distinguish between the continuum of sources specified by
\begin{equation}
 {\rm constant}\times\int{\rm d}^4x\bar{q}(x)e^{i\gamma_5\vec{\tau}\vec{\theta}}q(x)\,,~~~~~|\theta|\in[0,2\pi)\,.
\end{equation}
Therefore, the regular part of the vacuum susceptibility must be identical when measured as the response to any one of these sources, so that there is $\chi_R=\chi$ for all choices of $\vec{\theta}$. This is the content of the so-called ``Mexican hat'' potential, which is used in building effective models for QCD. The magnitude of $\chi$ depends on whether the chiral symmetry is dynamically broken and the strength of the interaction as measured with respect to the critical value required for DCSB, as discussed in Sec.~\ref{scalarsus} of this paper~\cite{PRC.79.035209}. When the symmetry is dynamically broken, then the Goldstone modes appear, by convention, in the pseudo-scalar--isovector channel, and thus the pole contributions appear in $\chi_5$ but not in the chiral susceptibility. It is valid to draw an analogy with the Weinberg sum rules~\cite{PLB.669.327--330,PRL.18.507}.
Eq.~(\ref{e28}) also provides a novel, model independent perspective on a mismatch between the evaluation of the pseudo-scalar vacuum susceptibility using either a two-point or a three-point sum rules. The two-point study in Ref.~\cite{PLB.136.273--278} produces the pion pole contribution, $\chi_5^\pi$, which is also the piece emphasized in Ref.~\cite{EPJA.16.291--297}, whereas a three-point method in Ref.~\cite{PRD.57.2847} isolates the regular piece, $\chi_5^R$, since a vacuum saturation ansatz is implemented in the derivation. Thus their analyses are not essentially in conflict, but emphasize different and independent pieces of the susceptibility, which can be distinguished. However, only the regular piece should be retained in a sum rules estimate of the pion-nucleon coupling constants~\cite{NPB.213.109--121}. We note in closing that other vacuum susceptibilities can be analyzed similarly.

\section{Susceptibilities and the chiral phase transition of QCD}\label{apps}
The chiral phase transition of QCD and a possible CEP on the QCD phase diagram have been drawing much attention on both experimental and theoretical sides; for example, beam energy scan (BES) program at the RHIC~\cite{JPG.38.124023,NPA.862.125--131,NPA.904.256c--263c,NPA.904.903c--906c}, and some theoretical studies~\cite{JHEP.09.073,PLB.643.46--54,Nature.443.675--678,JHEP.06.088,PLB.110.155--158,PRD.29.338,PRD.41.1610,
PRD.59.034010,PRD.75.085015,PhysRevLett.98.092301,PPN.39.1025--1032,PRD.79.014018,PhysRevLett.103.052003,PhysRevD.80.074029,PLB.702.438,PLB.718.1036,PPNP.67.200,PRD.90.034022}. To study the chiral phase transition of QCD, in the chiral limit the two-quark condensate can serve as an order parameter. Nevertheless, the current quark mass is small but not zero in the real world, so when studying the partial restoration of chiral symmetry at finite temperature and quark chemical potential, people also need to resort to some alternative order parameters. When there is a coexistence of different phases, one of the candidates is the pressure difference (the bag constant)~\cite{PRD.85.034031,EPJC.73.2612,JHEP.7.014,AoP.348.306}. As discussed above, various susceptibilities are the parameters that related to the linear responses of the system to the external fields, which can then characterize the properties of the system, as well as to serve as indicators of corresponding phase transitions. In this section, we focus on the applications of some QCD susceptibilities to the chiral phase transition of QCD.

In quantum field theory, the dynamic properties of a system are fully characterized by the generating functional, which corresponds to the partition function in thermodynamics. When the system is in a certain phase, the generating functional is usually analytic for some choice of parameters, such as the current masses of the fermions, the chemical potential, and the temperature. The generating functional often exhibits a non-analytic character while the phase transitions occur. So the location and characteristics of the chiral phase transition in the system can be determined by the behaviors of this quantity with respect to the corresponding parameters. In this case, a phase transition, in which the first order derivative of the generating functional with some of the parameters is discontinuous, is referred to as first order or discontinuous phase transition. Second order or continuous phase transition exhibits continuity in first order derivative and a discontinuity or infinity in second order derivative. Nevertheless, thanks to the non-perturbative properties of QCD in the low energy region, such discussions are difficult and complicated. Accordingly, researchers often resort to some effective models, for example, see Refs.~\cite{PhysRevD.90.073013,AoP.348.306,PRC.64.045202,AOP.322.701,PRD.75.034007,PRD.76.074023,PLB.662.26--32,PLB.696.58--67,PRD.86.105042,FBS.55.47--56,EPJC.74.2782}. In the following, we take the NJL model, which is successful as well as popular in chiral phase transition studies~\cite{NPA.504.668--684,PLB.249.386--390,PLB.506.267--274,PhysRevD.77.096001,PhysRevD.81.074005,PhysRevD.81.116005,PhysRevLett.106.142003,EPJC.73.2612,PhysRevD.85.074003,AOP.354.72,PhysRevD.91.036006}, as an example, and give some discussions on the applications of some susceptibilities~\cite{PhysRevD.88.114019}, similar discussions with in the framework of QED$_3$ can be found in Ref.~\cite{PhysRevD.90.036007}. For the details of the NJL model, please see Refs.~\cite{NPA.504.668--684,RMP.64.649,PR.247.221,PR.407.205}.

The Lagrangian of the NJL model is (here we still work in the Euclidean space, and take the number of flavors $N_f=2$ while the number of colors $N_c=3$)
\begin{equation}
\mathcal{L}_{\rm NJL}=\mathcal{L}_0+g\mathcal{L}_{\rm int}=\bar{q}(i\!\not\!\partial-m)q+g[(\bar{q}q)^2+(\bar{q}i\gamma_5\tau q)^2],
\end{equation}
where $g$ is a coupling constant with the dimension of mass$^{-2}$, and the flavor and color indices are suppressed. Now we introduce the definitions of four kinds of susceptibilities: the chiral susceptibility $\chi_s$, the quark number susceptibility $\chi_q$, the vector-scalar susceptibility $\chi_{vs}$, and another auxiliary susceptibility  $\chi_m$. For mathematical convenience, we first introduce these susceptibilities in the free quark gas case, where the interaction term in the Lagrangian is zero, $\mathcal{L}_{\rm int}=0$. Denoting them with the superscript $^{(0)}$, their definitions and expressions are then
\begin{eqnarray}
  \chi_s^{(0)} &\equiv& -\frac{\partial\langle\bar{q}q\rangle_f}{\partial m}=\frac{N_cN_f}{\pi^2}\int_0^\Lambda\bigg[\frac{m^2p^2\beta}{E_0^2}g(\mu)+\frac{p^4}{E_0^3}f(\mu)\bigg]{\rm d}p,\label{eq:chis} \\
  \chi_q^{(0)} &\equiv& \frac{\partial\langle q^{\dagger}q\rangle_f}{\partial \mu}=\frac{N_cN_f}{\pi^2}\int_0^\Lambda p^2\beta g(\mu){\rm d}p, \label{eq:chiq} \\
  \chi_{vs}^{(0)} &\equiv& \frac{\partial\langle\bar{q}q\rangle_f}{\partial \mu} =\frac{N_cN_f}{\pi^2}\int_0^\Lambda\frac{mp^2\beta}{E_0}h(\mu){\rm d}p, \label{eq:chivs} \\
  \chi_{m}^{(0)} &\equiv& -\frac{\partial\langle q^{\dagger}q\rangle_f}{\partial m}=\chi_{vs}^{(0)}, \label{eq:chim}
\end{eqnarray}
where $\Lambda$ is a three-momentum cutoff which is introduced to avoid the ultraviolet divergence in the NJL model, $\beta=1/T$, $E_0=\sqrt{m^2+p^2}$, $g(\mu)+h(\mu)=2n(\mu)(1-n(\mu))$, $g(\mu)-h(\mu)=2m(\mu)(1-m(\mu))$, $f(\mu)=1-n(\mu)-m(\mu)$, and
\begin{equation}\label{eqn:nm}
  n(m,p,\mu)=\frac{1}{1+\exp{[\beta(E_0-\mu)]}},
\end{equation}
\begin{equation}
  m(m,p,\mu)=\frac{1}{1+\exp{[\beta(E_0+\mu)]}},
\end{equation}
the subscript $f$ represents the free quark gas systems. It should be noted that $\chi_{m}^{(0)}$ and $\chi_{vs}^{(0)}$ have the same analytical expression, which is reasonable from the viewpoint of statistical mechanics,
\begin{equation}\label{eqn:mixed partial derivative}
\chi_{m}^{(0)}=\chi_{vs}^{(0)}=\frac{T}{V}\frac{\partial^2}{\partial m \partial \mu}\ln Z_{f},
\end{equation}
with $Z_{f}$ the QCD partition function in the free quark gas case.

In the interacting case, these susceptibilities are coupled with each other,
\begin{eqnarray}
 \chi_s&\equiv&-\frac{\partial\langle\bar{q}q\rangle}{\partial m}=\chi_{s}^{(0)}(\mu)(1+2g\chi_s)-\frac{g}{N_c}\chi_{vs}^{(0)}(\mu)\chi_m, \label{eq:chisc}\\
 \chi_q&\equiv&\frac{\partial\langle q^{\dagger}q\rangle}{\partial \mu}=2g\chi_{vs}^{(0)}(\mu)\chi_{vs}+ \chi_q^{(0)}(\mu)(1-\frac{g}{N_c}\chi_q),\label{eq:chiqc}\\
 \chi_{vs}&\equiv&\frac{\partial\langle\bar{q}q\rangle}{\partial \mu}=2g\chi_{s}^{(0)}(\mu)\chi_{vs}+\chi_{vs}^{(0)}(\mu)(1-\frac{g}{N_c}\chi_q) ,\label{eq:chivsc}\\
 \chi_{m}&\equiv&-\frac{\partial\langle q^{\dagger}q\rangle}{\partial m}=\chi_m^{(0)}(\mu)(1+2g\chi_s)-\frac{g}{N_c}\chi_q^{(0)}(\mu) \chi_m.\label{eq:chimc}
\end{eqnarray}

\begin{figure}[htb]
  \centering
  \includegraphics[width=0.7\textwidth]{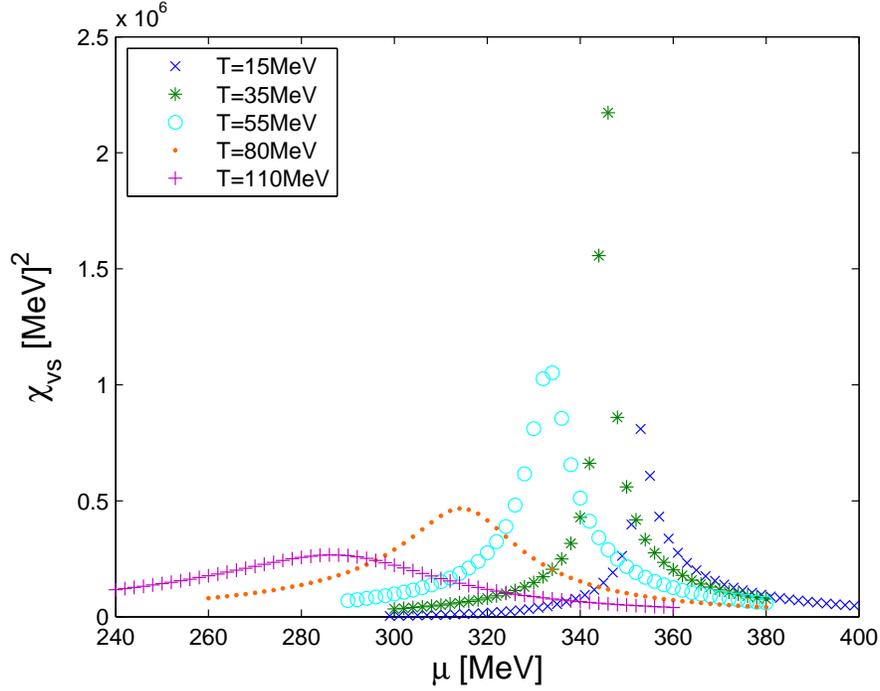}
  \caption{$\chi_{vs}$ at different $\mu$ and $T$, taken from Ref.~\cite{PhysRevD.88.114019}.}
  \label{fig:gra7chivsc}
\end{figure}

Using the iterative method, we can obtain the numerical results of these susceptibilities. For example, the vector-scalar susceptibility is the response of the effective quark mass ($M=m-2g\langle\bar{q}q\rangle$) to the chemical potential $\mu$, and its results are shown in Fig.~\ref{fig:gra7chivsc}. The parameters used in this section are taken from Ref.~\cite{PR.247.221}, namely, $m=5.5$~MeV, $g=5.074$~GeV$^{-2}$, and $\Lambda=631$~MeV.

\begin{figure}[htb]
\centering
\includegraphics[width=0.48\textwidth]{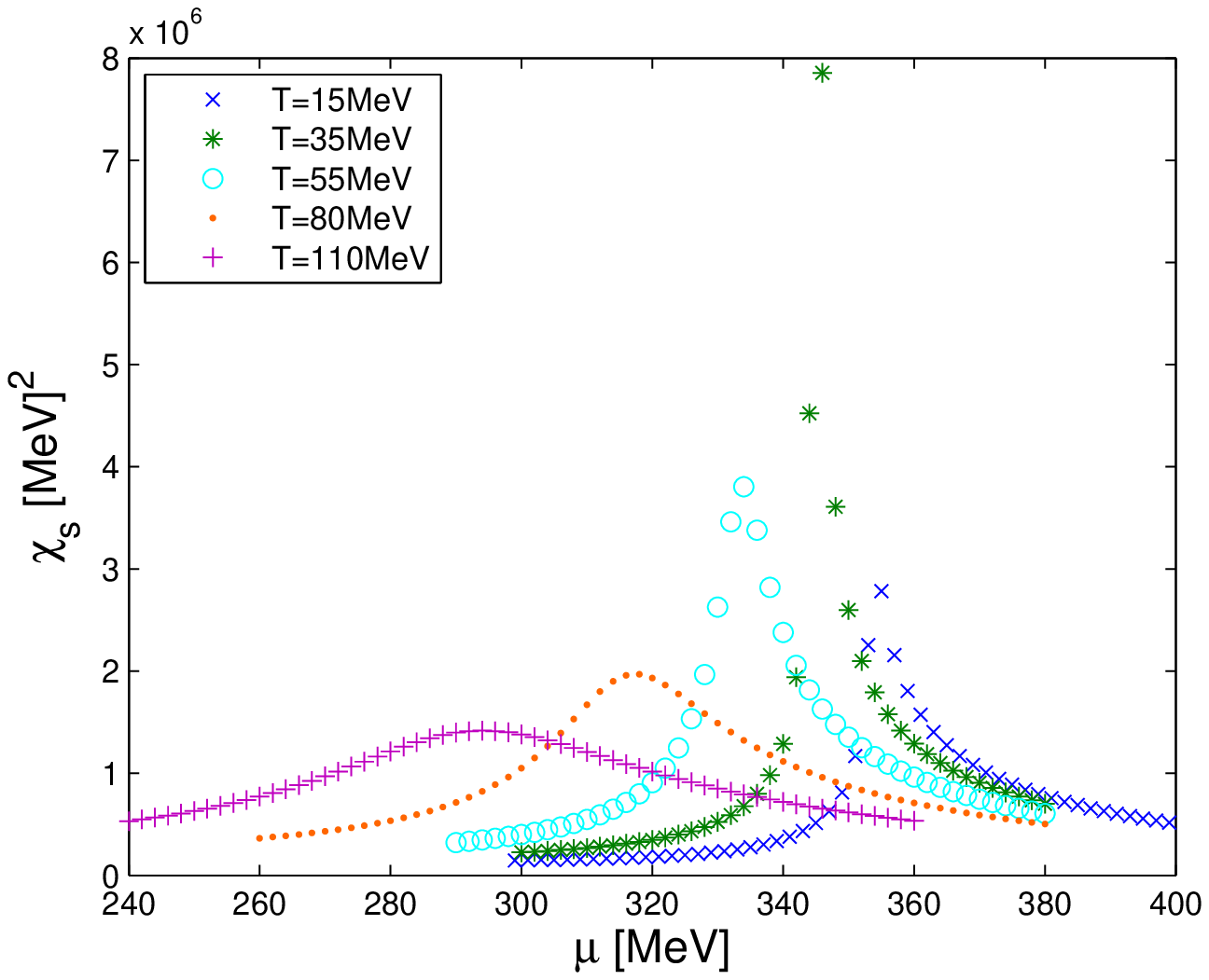}
\includegraphics[width=0.47\textwidth]{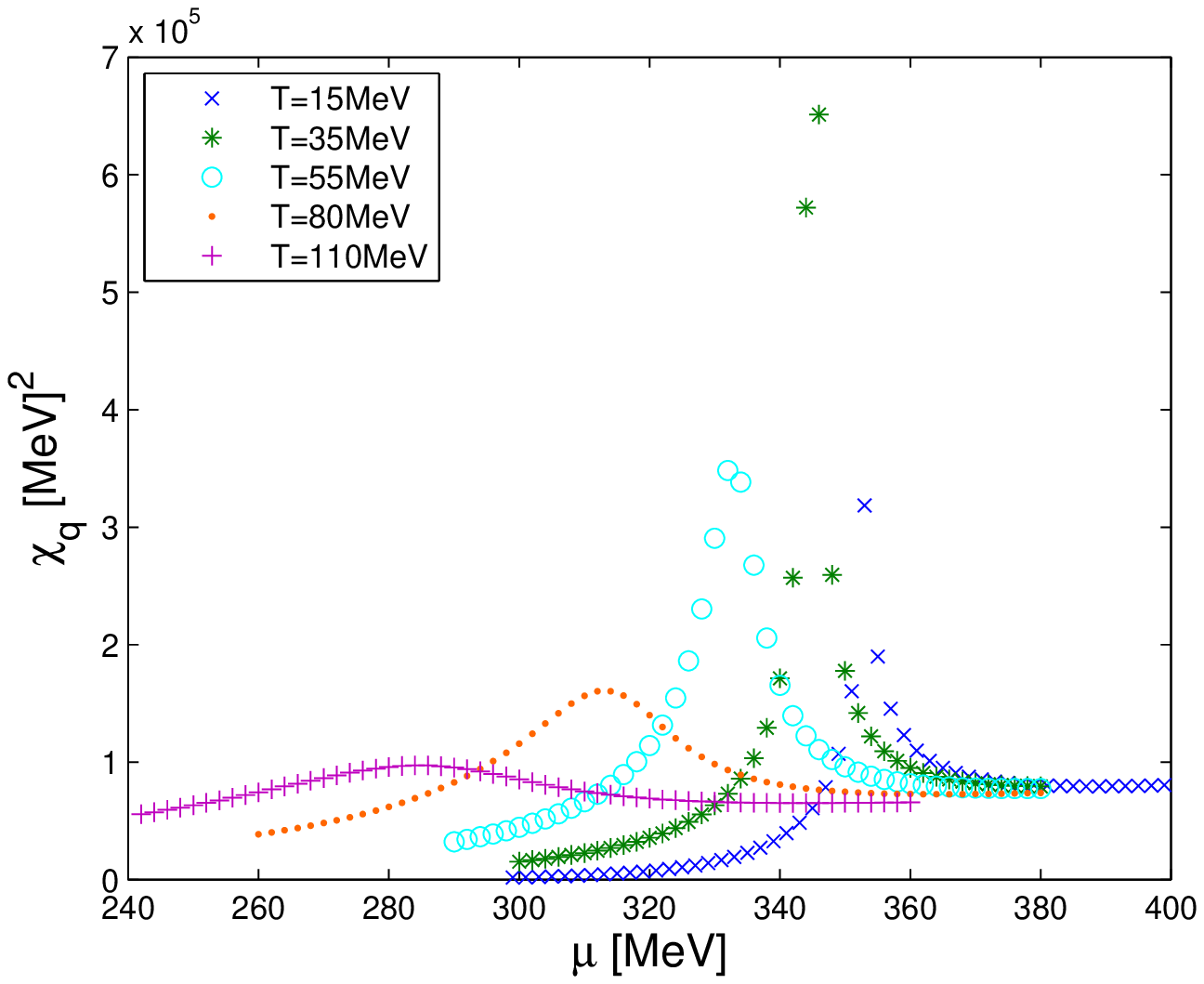}
\caption{\label{chisq} $\chi_{s}$ and $\chi_{q}$  at different $\mu$ and $T$, taken from Ref.~\cite{PhysRevD.88.114019}.}
\end{figure}

It can be seen from Fig.~\ref{fig:gra7chivsc} that, when $T$ is smaller than a critical value $T_c=35$~MeV, there always exists a convergent discontinuity of $\chi_{vs}$, corresponding to a first order phase transition; when $T=T_c$, $\chi_{vs}$ displays a sharp and narrow divergent peak, which implies a second-order phase transition, or in other words, here is a CEP on the phase diagram; when $T>T_c$, the discontinuity disappears and a rather broad peak of finite height is shown, corresponding to the crossover region. Comparing this with the results of the chiral susceptibility $\chi_s$ and the quark number susceptibility $\chi_q$, shown in Fig.~\ref{chisq}, we can conclude that at low temperature, the first order phase transition occurs at almost the same chemical potential; while in the crossover region, if we pick the peak of these susceptibilities as the artificial critical points, we see that they tend to occur at different chemical potentials as the temperature increases.

\begin{figure}[htb]
  \centering
  \includegraphics[width=0.48\textwidth]{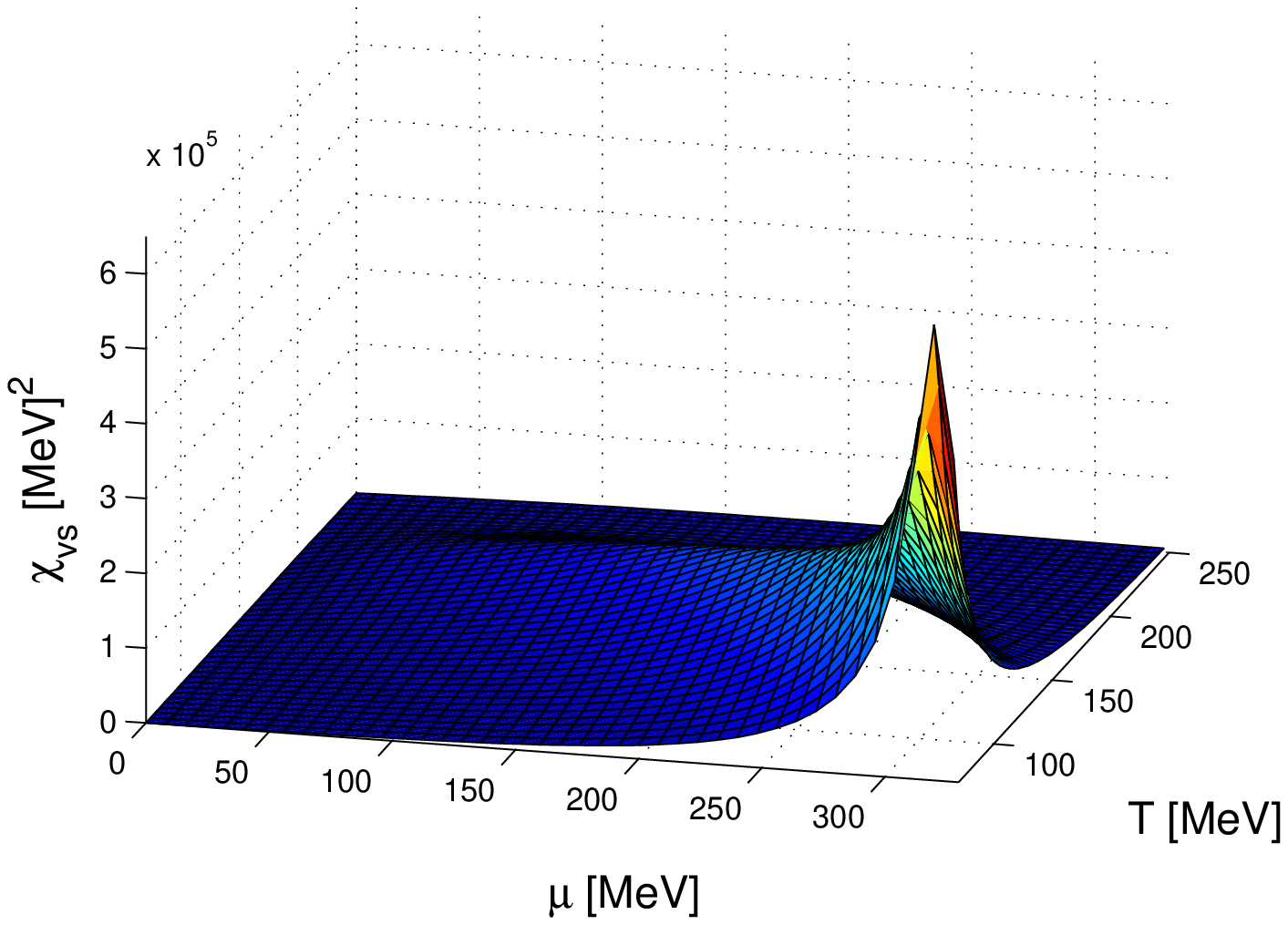}
  \includegraphics[width=0.48\textwidth]{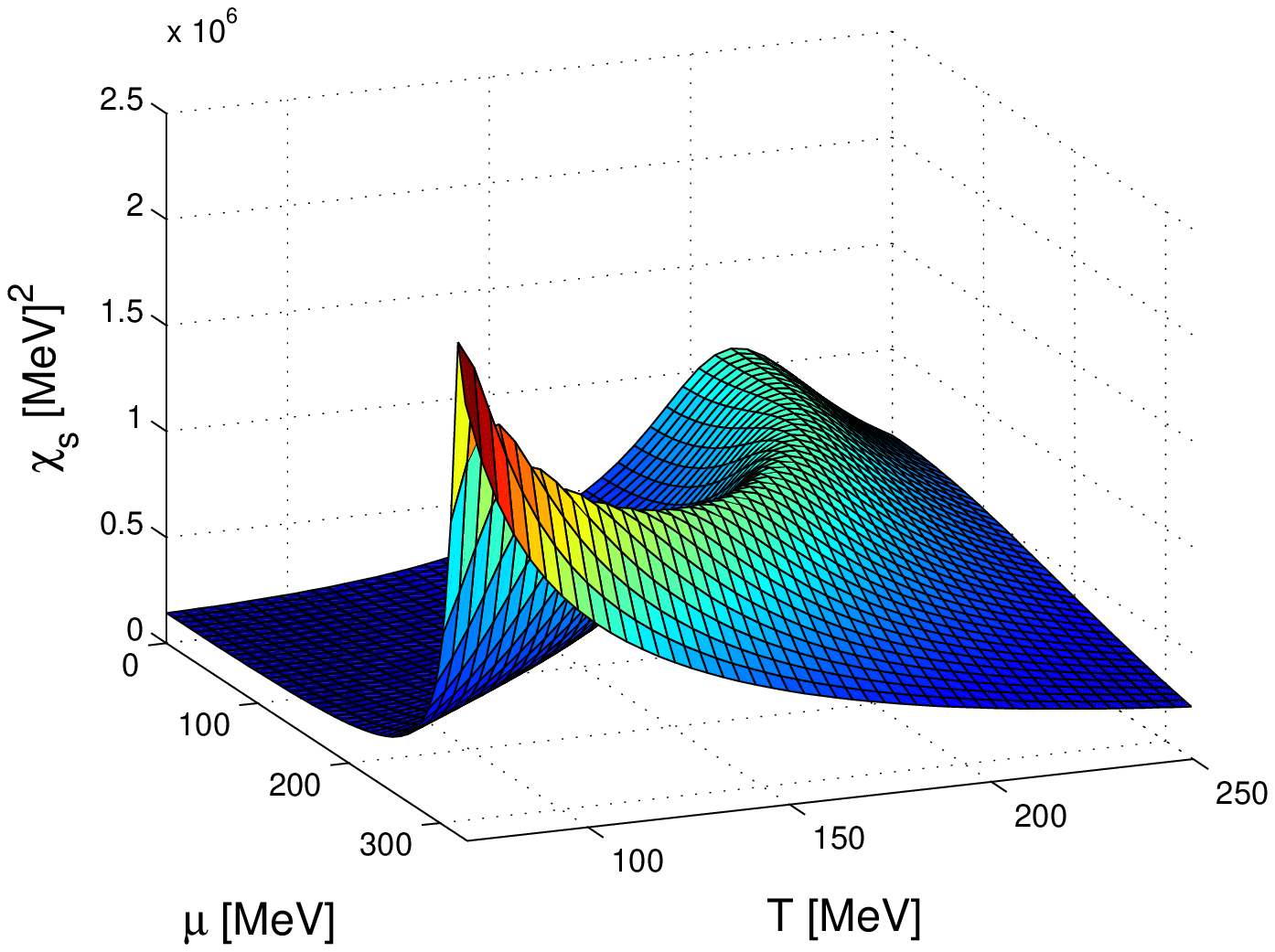}
  \caption{$\chi_{vs}$ and $\chi_{s}$ in the crossover region, taken from Ref.~\cite{PhysRevD.88.114019}.}\label{chivss3d}
\end{figure}

\begin{figure}[htb]
  \centering
  \includegraphics[width=0.48\textwidth]{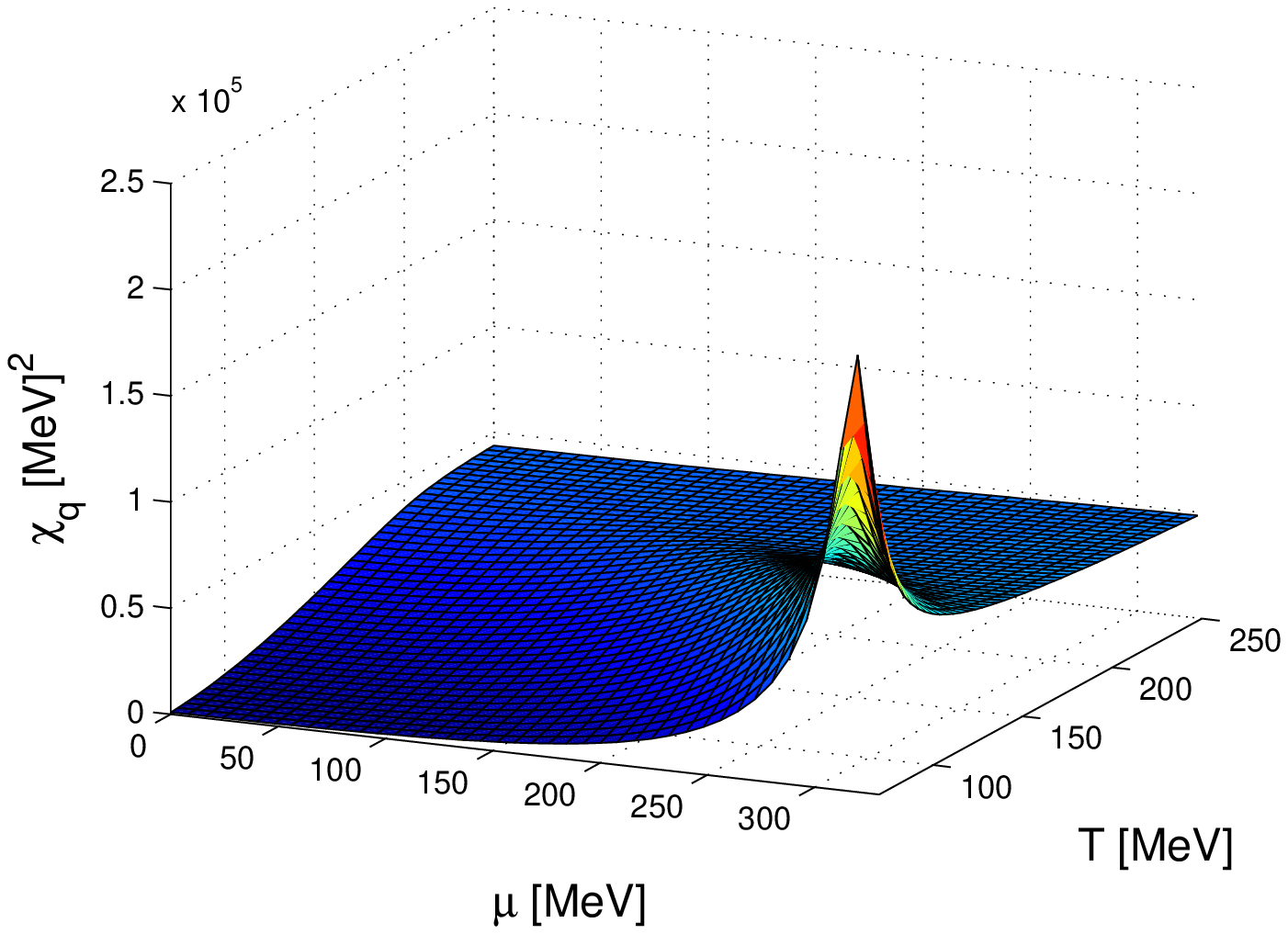}
  \includegraphics[width=0.48\textwidth]{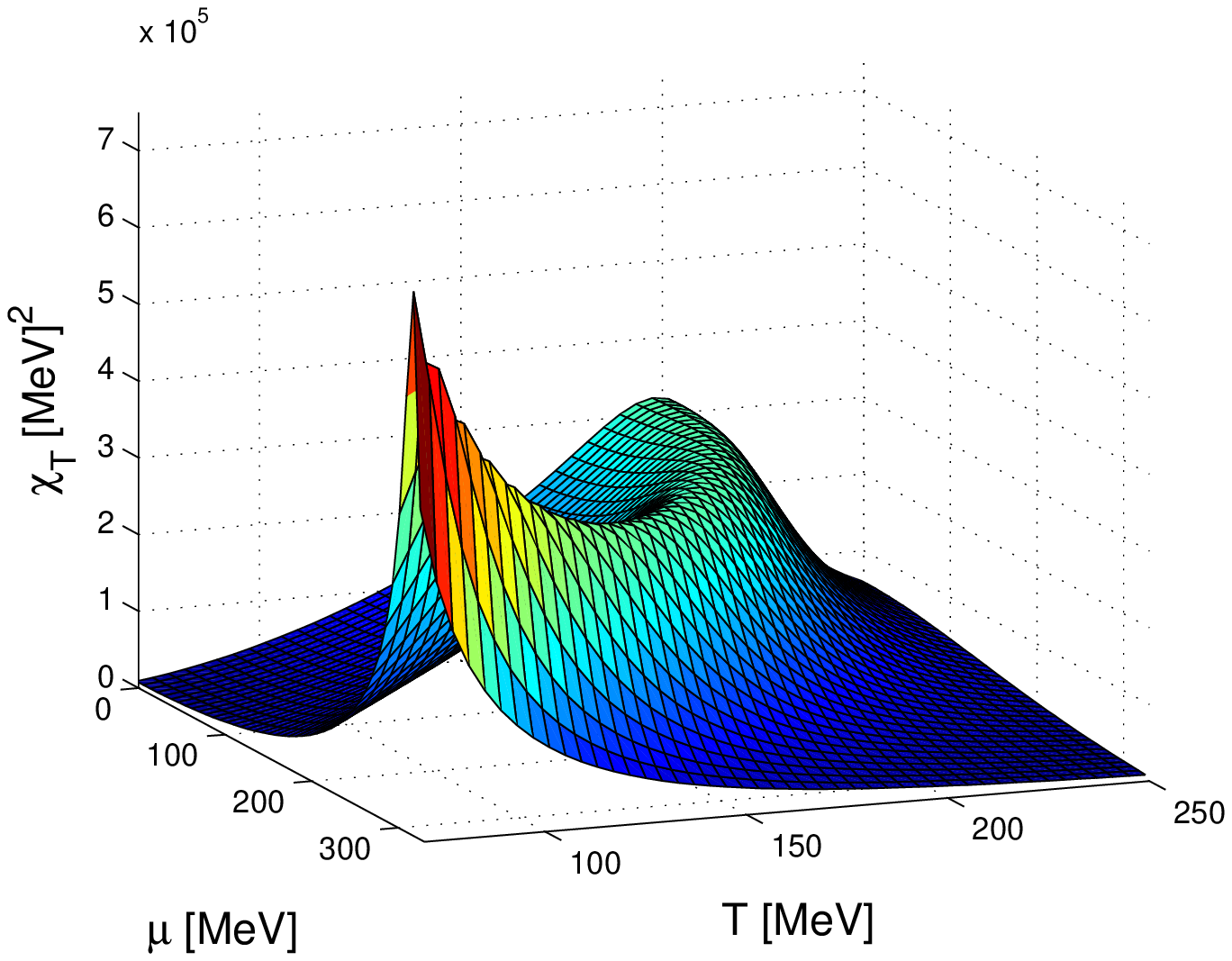}
  \caption{$\chi_{q}$ and $\chi_T$ in the crossover region, taken from Ref.~\cite{PhysRevD.88.114019}.}\label{chiqT3d}
\end{figure}

The calculated results of $\chi_{vs}$, $\chi_s$, and $\chi_q$ in the crossover region are shown in Figs.~\ref{chivss3d} and~\ref{chiqT3d}, respectively. We can find that they exhibit different behaviors: the chiral susceptibility $\chi_s$ exhibits an obvious band, so it is convincing to define the peak of $\chi_s$ as the artificial critical points; in the high $T$ and/or low $\mu$ region, the vector-scalar susceptibility $\chi_{vs}$ tends to vanish; while the global shape of the quark number susceptibility $\chi_q$ is just similar to the ones of $\chi_{s}$ and $\chi_{vs}$, but it is non-vanishing in the high $T$ and/or high $\mu$ region whose behavior is closely linked to the quark number density. Therefore, $\chi_q$ can not describe the crossover property well in the high $T$ and/or high $\mu$ region.

It is very interesting and meaningful to compare the results of these susceptibilities with that of the thermal susceptibility $\chi_T=\partial\langle\bar{q}q\rangle/\partial T$. For mathematical convenience, we also define $\chi_n=\partial\langle q^{\dagger}q\rangle/\partial T$. By the same process employed above, we obtain a set of coupled equations for $\chi_T$ and $\chi_n$ as
\begin{gather}
\begin{split}
\chi_T\equiv&\frac{\partial\langle\bar{q}q\rangle}{\partial T}=\frac{\partial\langle\bar{q}q\rangle}{\partial M}\frac{\partial M}{\partial T}+\frac{\partial\langle\bar{q}q\rangle}{\partial \mu}\frac{\partial \mu}{\partial T}+\big(\frac{\partial\langle\bar{q}q\rangle}{\partial T}\big)_{M,\mu}\\
 =&2g\chi_s^{(0)}(\mu)\chi_T-\frac{g}{N_c}\chi_{vs}^{(0)}(\mu)\chi_n+M\beta\chi_q^{(0)}(\mu)-\mu\beta\chi_m^{(0)}(\mu), \label{eq:chiT}
\end{split}\\
\begin{split}
 \hspace{0mm}\chi_n\equiv&\frac{\partial\langle q^{\dagger}q\rangle}{\partial T}=\frac{\partial\langle q^{\dagger}q\rangle}{\partial M}\frac{\partial M}{\partial T}+\frac{\partial\langle q^{\dagger}q\rangle}{\partial \mu}\frac{\partial \mu}{\partial T}+\big(\frac{\partial\langle q^{\dagger}q\rangle}{\partial T}\big)_{M,\mu}\\
 =&2g\chi_{m}^{(0)}(\mu)\chi_{T}- \frac{g}{N_c}\chi_q^{(0)}(\mu)\chi_n-\mu\beta\chi_q^{(0)}(\mu)+\frac{N_cN_f}{\pi^2}\int_0^\Lambda p^2E\beta^2h(\mu){\rm d}p,\label{eq:chin}
\end{split}
\end{gather}
with $E_=\sqrt{M^2+p^2}$. The behavior of $\chi_T$ in the crossover region is shown in Fig.~\ref{chiqT3d}, which is very similar to that of $\chi_s$.

\begin{figure}[htb]
  \centering
  \includegraphics[width=0.7\textwidth]{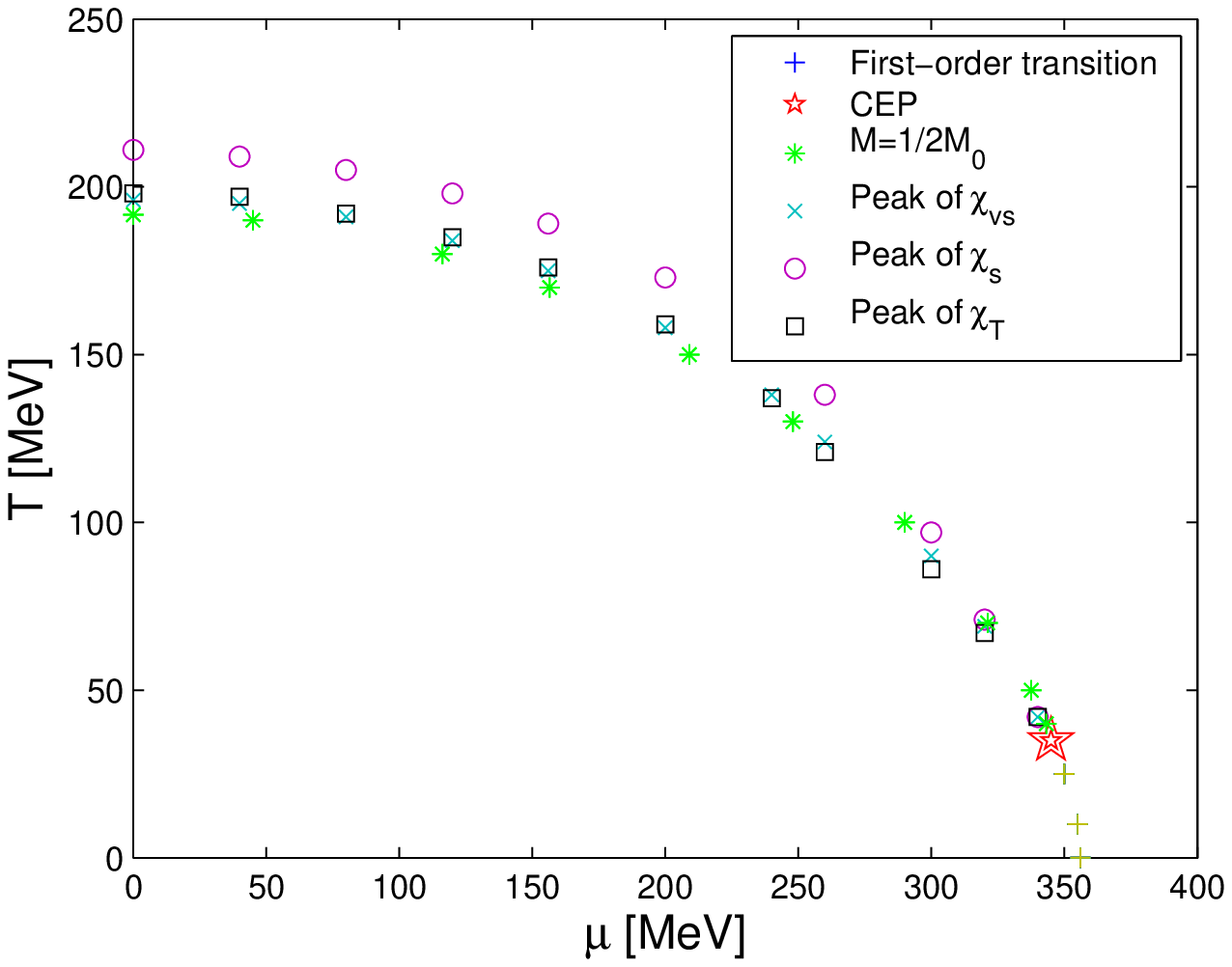}
  \caption{Phase diagram obtained according to different susceptibilities, taken from Ref.~\cite{PhysRevD.88.114019}.}
  \label{fig:phasedia}
\end{figure}

\begin{figure}[htb]
  \centering
  \includegraphics[width=0.7\textwidth]{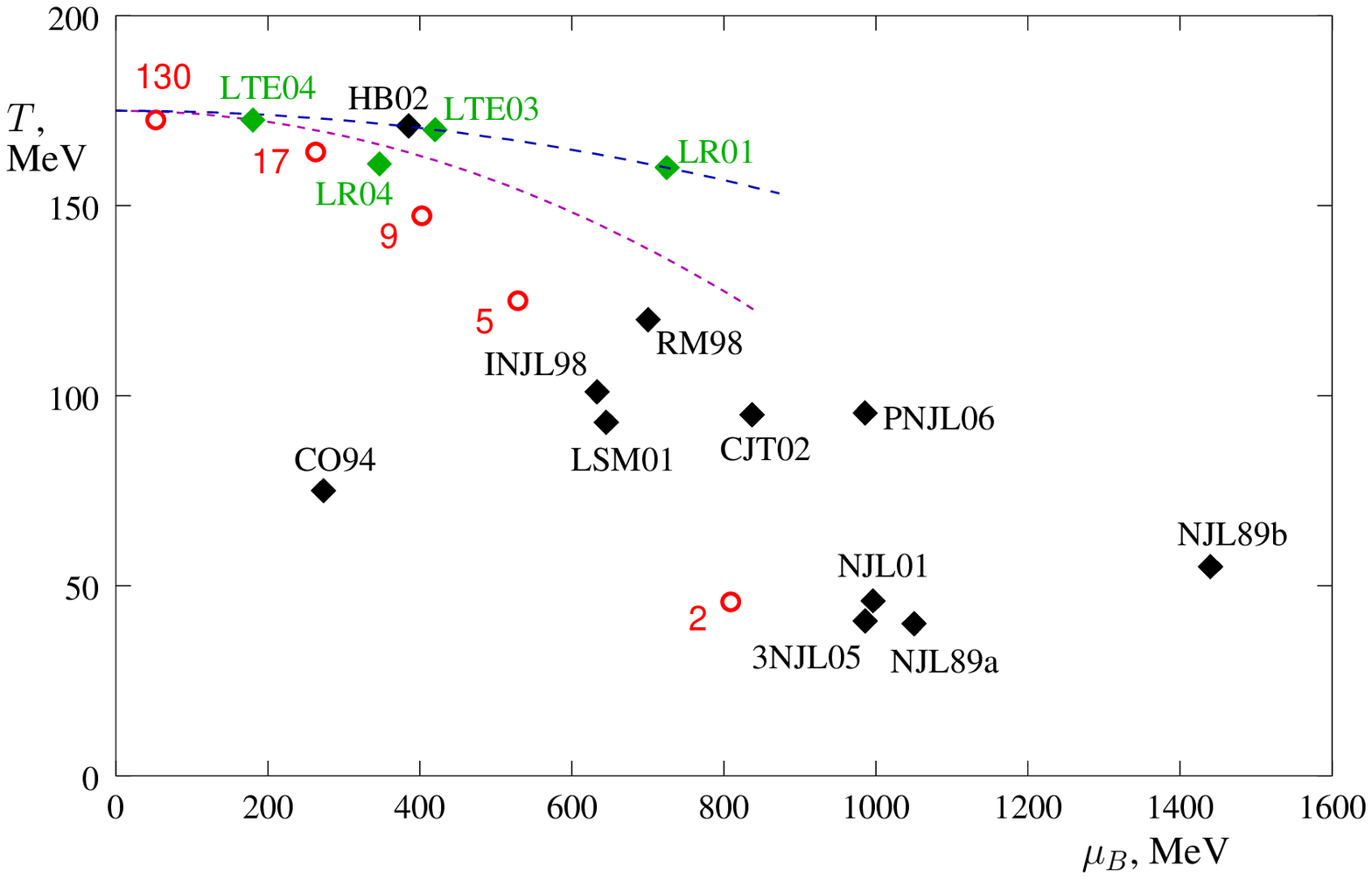}
  \caption{Comparison of model and Lattice QCD predictions for the location of the possible CEP on the QCD phase diagram, taken from Ref.~\cite{PoS.06.024} (please refer to this paper for more details).}
  \label{ceps}
\end{figure}

Now we compare the results of different susceptibilities in a chiral phase diagram of QCD, as presented in Fig.~\ref{fig:phasedia}. The corresponding lines in the crossover region are determined by the peaks $\chi_{vs}$, $\chi_T$, and $\chi_s$. Here it should be stressed that, actually there is a very large number of studies investigating the phase diagram together with the CEP of QCD, and in various models~\cite{PoS.06.024,EPJC.73.2612,JHEP.7.014,FBS.55.47--56,IJMPA.20.4387,PhysRevLett.106.172301,AIP.1355.101,PhysRevD.85.091901,1411.4953,PhysRevD.91.034017,PhysRevD.91.056003} as well as functional and Lattice QCD studies~\cite{JHEP.09.073,PhysRevD.77.014508,JHEP.04.01,JHEP.02.044,PhysRevD.85.054503,PhysRevLett.110.172001,PhysRevLett.113.082001,PhysRevLett.113.152002} the results vary quite widely, most of them find a higher value for $T_c$ and a lower value for $\mu_c$ of the CEP we plotted here, {as summarised in Fig.~\ref{ceps} (here $\mu_B=3\mu$), which is taken from Fig.~4 of Ref.~\cite{PoS.06.024}. The points in Fig.~\ref{ceps} represent effective model and Lattice QCD predictions, the two dashed lines are parabolas with slopes corresponding to Lattice QCD calculations, and the red circles are locations of the freeze-out points for heavy ion collisions at corresponding center of mass energies per nucleon (indicated by labels in GeV). It is shown clearly that, different models or parameter sets give quite different predictions (please refer to Ref.~\cite{PoS.06.024} for more details of these points and lines)}. The parameters we adopted here ({NJL89a in Fig.~\ref{ceps}}) are just for the convenience of showing the qualitative results of different susceptibilities more explicitly. For a recent parameter-independent attempt of such studies, please see Ref.~\cite{IJMPcui}.

We see from Fig.~\ref{fig:phasedia} that these susceptibilities split when the system tends to lower quark chemical potential and/or higher temperature. As we discussed above, up to now there is no exact order parameter for studies of QCD chiral phase transition when beyond the chiral limit, and different susceptibilities would characterize different physical properties of the system. Hence, related topics deserve further discussions.

\section{Summary}
In the Dyson-Schwinger equations (DSEs) framework of QCD, we summarize studies of various vacuum susceptibilities together with some of their applications to the chiral phase transition of QCD. In Section~\ref{intro}, we give a brief introduction of the vacuum susceptibilities together with their related applications. Then a general derivation of the vacuum susceptibility using the QCD sum rules external field formula is explicated in the following Section~\ref{sumrules}, as well as a model independent expression, which is expressed with the quark propagator (two-point Green function) and the corresponding vertex function (three-point Green function). In the next Section~\ref{calsus}, we review the calculations of the quark propagator and the vertex function under the framework of DSEs-BSEs first, and then use these theoretical methods and results to discuss the calculations of the tensor, the vector, the axis-vector, the scalar, and the pseudo-scalar vacuum susceptibilities further. For the tensor vacuum susceptibility, we introduce the calculation results of the Maris-Tandy gluon propagator model using the rainbow-ladder approximation of DSEs. For the vector and axial-vector vacuum susceptibilities, we present the model independent results obtained with the help of the vector and axial-vector Ward-Takahashi identities: the vector vacuum susceptibility is strictly $0$, while the axial-vector vacuum susceptibility equals $3/4$ of the square of the pion decay constant. For the scalar vacuum susceptibility, we show the results using a Gauss model (infrared part of the Maris-Tandy model) gluon propagator with the Ball-Chiu vertex ansatz, and comparisons with those from the bare vertex (the rainbow ladder truncation) are also given. We also show a straightforward explanation of a mismatch between extant estimates of the pseudo-scalar vacuum susceptibility. These works promoted our understandings of the five kinds of vacuum susceptibilities. In a word, by this time only the scalar and the tenser vacuum susceptibilities are still model-dependent, and to choose a more reliable model framework then becomes very important. With the help of an effective model of QCD then, in Section~\ref{apps} we show the applications of some susceptibilities to the QCD chiral phase transition as well as some discussions of the results.

Here it also needs to be pointed out that, in the QCD sum rules external field approach, all kinds of vacuum condensates and vacuum susceptibilities are introduced as independent phenomenological parameters, and their values are determined by fitting the theoretical predictions from QCD sum rules to the experimental results. The authors of Ref.~\cite{IJMPA.23.1507--1520} proposed for the first time that the four-quark condensate may relate to the corresponding vacuum susceptibility. If this conclusion is correct, the QCD sum rules external field formula can then introduce less independent parameters, which will certainly increase its {predictiveness}. Last but not least, it should be noted that up to now people still do not know how to rigorously define some of the vacuum condensates (including but not limited to the four-quark condensate) from the first principle of QCD. There is no doubt that, the issues related to the four-quark condensate and the relations between the four-quark condensate and the corresponding vacuum susceptibility are worth further studying. And of course, so are many discussions in this paper.

\section*{Acknowledgments}
This work is supported in part by the National Natural Science Foundation of China (under Grant 11275097, 11475085, and 11247219), the Jiangsu Planned Projects for Postdoctoral Research Funds (under Grant No. 1402006C), and the National Natural Science Foundation of Jiangsu Province of China (under Grant BK20130078).

\section*{References}
\bibliography{AOP70267}

\end{document}